\documentclass[10pt,conference,english]{IEEEtran}
\usepackage[centertags]{amsmath}
\usepackage{amsfonts}
\usepackage{amssymb}
\usepackage{epsfig}
\usepackage{amsmath,  amssymb}
\usepackage{graphicx}
\usepackage[centertags]{amsmath}
\usepackage{clrscode}
\usepackage{cite}
\usepackage{fullpage}
\usepackage{amsmath,amssymb,mathrsfs,pseudocode}
\usepackage{color}
\usepackage[normalem]{ulem}
\usepackage{psfrag}
\usepackage{url}
\usepackage{babel}
\usepackage{authblk}
\usepackage[linesnumbered,ruled]{algorithm2e}
\usepackage[margin=18.5mm]{geometry}
\usepackage{xspace}% essential for newcommand endings
\usepackage{multicol}
\usepackage{booktabs} % nice table objects
\usepackage{multirow}
\usepackage[x11names,dvipsnames,table]{xcolor}

\usepackage{subfigure}

\usepackage{stackrel}
\usepackage{extarrows}

\newcommand{\opertype}[1]{\begin{minipage}{29mm}\centering\vspace{1mm} #1\vspace{1mm}\end{m
inipage}}

\newtheorem{theorem}{Theorem}%[section]
\newtheorem{lemma}{Lemma}

\newtheorem{proposition}{Proposition}%[section]
\newtheorem{example}{Example}%[section]
\newtheorem{definition}{Definition}%[section]
\usepackage{xspace}
\usepackage{bbm}
\catcode`@=11
\chardef\mathlig@atcode\count255

\def\actively#1#2{\begingroup\uccode`\~=`#2\relax\uppercase{\endgroup#1~}}
\def\mathlig@gobble{\afterassignment\mathlig@next@cmd\let\mathlig@next= }
\def\mathlig@delim{\mathlig@delim}
\def\mathlig@defcs#1{\expandafter\def\csname#1\endcsname}
\def\mathlig@let@cs#1#2{\expandafter\let\expandafter#1\csname#2\endcsname}
\def\mathlig@appendcs#1#2{\expandafter\edef\csname#1\endcsname{\csname#1\endcsname#2}}

\def\mathlig#1#2{\mathlig@checklig#1\mathlig@end\mathlig@defcs{mathlig@back@#1}{#2}\ignorespaces}

\def\mathlig@checklig#1#2\mathlig@end{%
 \expandafter\ifx\csname mathlig@forw@#1\endcsname\relax
 \expandafter\mathchardef\csname mathlig@back@#1\endcsname=\mathcode`#1%
 \mathcode`#1"8000\actively\def#1{\csname mathlig@look@#1\endcsname}%
 \mathlig@dolig#1\mathlig@delim
\fi
\mathlig@checksuffix#1#2\mathlig@end
}

\def\mathlig@checksuffix#1#2\mathlig@end{%
\ifx\mathlig@delim#2\mathlig@delim\relax\else\mathlig@checksuffix@{#1}#2\mathlig@end\fi
}
\def\mathlig@checksuffix@#1#2#3\mathlig@end{%
\expandafter\ifx\csname mathlig@forw@#1#2\endcsname\relax\mathlig@dosuffix{#1}{#2}\fi
\mathlig@checksuffix{#1#2}#3\mathlig@end
}

\def\mathlig@dosuffix#1#2{%
\mathlig@appendcs{mathlig@toks@#1}{#2}%
\mathlig@dolig{#1}{#2}\mathlig@delim
}

\def\mathlig@dolig#1#2\mathlig@delim{%
 \mathlig@defcs{mathlig@look@#1#2}{%
 \mathlig@let@cs\mathlig@next{mathlig@forw@#1#2}\futurelet\mathlig@next@tok\mathlig@next}%
 \mathlig@defcs{mathlig@forw@#1#2}{%
  \mathlig@let@cs\mathlig@next{mathlig@back@#1#2}%
  \mathlig@let@cs\checker{mathlig@chck@#1#2}%
  \mathlig@let@cs\mathligtoks{mathlig@toks@#1#2}%
  \expandafter\ifx\expandafter\mathlig@delim\mathligtoks\mathlig@delim\relax\else
  \expandafter\checker\mathligtoks\mathlig@delim\fi
  \mathlig@next
 }%
 \mathlig@defcs{mathlig@toks@#1#2}{}%
 \mathlig@defcs{mathlig@chck@#1#2}##1##2\mathlig@delim{%
  \ifx\mathlig@next@tok##1%
   \mathlig@let@cs\mathlig@next@cmd{mathlig@look@#1#2##1}\let\mathlig@next\mathlig@gobble
  \fi 
  \ifx\mathlig@delim##2\mathlig@delim\relax\else
   \csname mathlig@chck@#1#2\endcsname##2\mathlig@delim
  \fi
 }%
 \ifx\mathlig@delim#2\mathlig@delim\else
  \mathlig@defcs{mathlig@back@#1#2}{\csname mathlig@back@#1\endcsname #2}%
 \fi
}%

\catcode`@\mathlig@atcode

\newcommand{\muspace}{\mspace{1mu}}

\DeclareRobustCommand{\scond}{\mathchoice{\muspace\vert\muspace}{\vert}{\vert}{\vert}}
\mathlig{|}{\scond}

\DeclareRobustCommand{\discint}{\mathchoice{\mspace{-1.5mu}:\mspace{-1.5mu}}{\mspace{-1.5mu}:\mspace{-1.5mu}}{:}{:}}
\mathlig{::}{\discint}

\newcommand{\Ac}{\mathcal{A}}
\newcommand{\Bc}{\mathcal{B}}
\newcommand{\Cc}{\mathcal{C}}

\newcommand{\Dc}{\mathcal{D}}

\newcommand{\Ic}{\mathcal{I}}

\newcommand{\Kc}{\mathcal{K}}

\newcommand{\Bcal}{\mathcal{B}}

\newcommand{\Gcal}{\mathcal{G}}

\newcommand{\Kcal}{\mathcal{K}}

\newcommand{\Xcal}{\mathcal{X}}

\newcommand{\Cr}{\mathscr{C}}

\newcommand{\Rr}{\mathscr{R}}

\newcommand{\Cv}{{\bf C}}
\newcommand{\Dv}{{\bf D}}

\newcommand{\Duv}{{\bf \underline D}}

\newcommand{\Xv}{{\bf X}}
\newcommand{\Pv}{{\bf P}}

\newcommand{\Yv}{{\bf Y}}

\newcommand{\Rv}{{\bf R}}

\newcommand{\tv}{{\bf t}}
\newcommand{\xv}{{\bf x}}
\newcommand{\yv}{{\bf y}}

\newcommand{\uv}{{\bf u}}

\newcommand{\rv}{{\bf r}}

\newcommand{\Xh}{{\hat{X}}}

\def\b{\beta}

\let\P\relax
\DeclareMathOperator\P{\textsf{P}}

\DeclareMathOperator\R{\textsf{R}}
\def\textiid{i.i.d.\@\xspace}
\newcommand\iid{\ifmmode\text{ i.i.d. } \else \textiid \fi}

\def\clap#1{\hbox to 0pt{\hss#1\hss}}
\def\mathclap{\mathpalette\mathclapinternal}
\def\mathclapinternal#1#2{%
  \clap{$\mathsurround=0pt#1{#2}$}}

\let\oldstackrel\stackrel
\renewcommand{\stackrel}[2]{\oldstackrel{\mathclap{#1}}{#2}}

\newcommand{\ntouch}[1]{\overline{{#1}}}

\begin{document}
\title{Generalized Alignment Chain: Improved Converse Results for Index Coding\vspace{-2mm}}
%\title{The single-uniprior index-coding problem: The single-sender case and the multi-sender extension\vspace{-2mm}}

\author{Yucheng Liu and Parastoo Sadeghi\\\vspace{-0mm}
Research School of Electrical,
Energy and Materials Engineering \\Australian National University, Canberra, ACT, 2601, Australia\\
%$^{*}$Department of Electrical and Computer Engineering, University of California, San Diego, CA 92093, USA\\
Emails:  \{yucheng.liu, parastoo.sadeghi\}@anu.edu.au}

\maketitle

\begin{abstract}

In this paper, we study the information-theoretic converse for the index coding problem. We generalize the definition for the alignment chain, introduced by Maleki et al., to capture more flexible relations among interfering messages at each receiver. Based on this, we derive improved converse results for the single-server index coding problem. Compared to the maximum acyclic induced subgraph (MAIS) bound, the new bounds are always as tight and can strictly outperform the MAIS bound. They can also be useful for large problems, where the generally tighter polymatroidal bound is computationally impractical. We then extend these new bounds to the multi-server index coding problem. We also present a separate, but related result where we identify a smaller single-server index coding instance, compared to those identified in the literature, for which non-Shannon-type inequalities are necessary to give a tighter converse.

\end{abstract}

\vspace{-1mm}
\section{Introduction}\label{sec:intro}

Index coding, introduced by Birk and Kol in \cite{Birk--Kol1998}, investigates the broadcast rate of $n$ messages from a server to multiple receivers with side information. %Each receiver wants to decode a unique message and has prior knowledge of some other messages as its side information. %During the past two decades, there has been substantial progress in theory and applications of index coding. See \cite{arbabjolfaei2018fundamentals} and the references therein. However, in general, the index coding problem remains open. 
Despite the substantial progress achieved during the past two decades, the index coding problem remains open in general.  See \cite{arbabjolfaei2018fundamentals} and the references therein.

In contrast to the single-server centralized index coding (CIC) problem, in the more general {distributed} index coding (DIC) problem, different subsets of the messages are stored at multiple servers. Such communication model has clear applications for practical circumstances where the information is geographically distributed over multiple locations. 
See \cite{Ong--Ho--Lim2016,liu2018capacity,li2019cooperative} and the references therein. 

In this paper, we study the information-theoretic converse for both CIC and DIC problems. 
For the CIC problem, the maximum acyclic induced subgraph (MAIS) bound was proposed in \cite{bar2011index}, and the polymatroidal (PM) bound %based on the polymatroidal axioms of the entropy function 
was presented in \cite{blasiak2011lexicographic, Arbabjolfaei--Bandemer--Kim--Sasoglu--Wang2013}. Both bounds have been extended to the DIC problem \cite{liu2018capacity,li2019cooperative}. 
The PM bound is generally tighter than the MAIS bound. However, it has a much higher computational complexity, which can be forbidding for large problems. %a large number of messages. 

%Therefore, it is of interest to find a middle ground between the easily computable, yet looser MAIS bound and the high-complexity, yet tighter PM bound. 
Therefore, it is of interest to find bounds that are strictly tighter than the MAIS bound, and at the same time, not as computationally intensive as the PM bound. 
The internal conflict bound for the CIC problem, introduced in \cite{maleki2014index,jafar2014topological} based on the \emph{alignment chain} model, can sometimes be useful. However, it does not subsume the MAIS bound in general. %and these two bounds can outperform each other for different CIC instances. 

In this paper, we first generalize the internal conflict bound for the CIC problem by extending the alignment chain model. 
%The new converse results, which are of closed-form given the generalized alignment chains, uniformly outperform the internal conflict bound and the MAIS bound. We show by examples that such relationships are sometimes strict. 
We prove that the new converse results are no looser than the internal conflict bound and the MAIS bound. We show by examples that they can sometimes be strictly tighter. 
We then generalize these results to the DIC problem. % (Thms. \ref{thm:wac:dic}, \ref{thm:gwac:dic} in Section \ref{sec:gwac:dic})
%, and show their efficacy via another example. 
%Finally in Section \ref{sec:nonshannon}, w
We also present a separate result. That is, we identify a smaller CIC problem in terms of the number of messages, compared to those previously identified in \cite{sun2015index},\cite[Section 5.3]{arbabjolfaei2018fundamentals}, 
where Shannon-type inequalities are insufficient to give a tight converse. 

{\it Notation:} For non-negative integers $a$ and $b$, $[a]$ denotes the set $\{1,2,\cdots,a\}$, and $[a:b]$ denotes the set $\{a,a+1,\cdots,b\}$. If $a>b$, $[a:b]=\emptyset$. For a set $S$, $|S|$ denotes its cardinality.

\section{System Model}\label{sec:model}

Assume that there are $n$ messages, $x_i \in \{0,1\}^{u_i}, i \in [n]$, where $u_i$ is the length of binary message $x_i$.  
For brevity, when we say message $i$, we mean message $x_i$. 
Let $X_i$ be the random variable corresponding to $x_i$. We assume that $X_1, \ldots, X_n$ are independent and uniformly distributed. 
For any $S\subseteq [n]$, set $S^c \doteq [n]\setminus S$, $\xv_S \doteq (x_i,i\in S)$, and $\Xv_S \doteq (X_i,i\in S)$. 
%We use the shorthand notation $\xv_S$ and $\Xv_S$ to denote the collection of messages and message random variables, whose index is in $S$, respectively. 
By convention, $\xv_{\emptyset} = \Xv_{\emptyset} = \emptyset$. 

There are $n$ receivers, where receiver $i \in [n]$ wishes to obtain $x_i$ and knows $\xv_{A_i}$ as side information for some $A_i \subseteq [n]\setminus \{i\}$. The set of indices of \emph{interfering messages} at receiver $i$ is denoted by the set $B_i=(A_i\cup \{i\})^c$. 

To avoid redundancy, 
%in the rest of this section, 
we describe the remaining system model for the DIC problem only, in which there are $2^n-1$ servers, each containing a unique nonempty subset of the $n$ messages. The server indexed by $J\in N$ contains messages $\xv_J$, where $N \doteq \{J \subseteq [n]: J \neq \emptyset\}$. Every server is connected to all receivers via its own noiseless broadcast channel with finite link capacity $C_J\ge 0$. Clearly, the CIC problem is a special case of the DIC problem with $C_{[n]}=1$ (normalized) and $C_J=0$ otherwise. Let $y_J\in \{0,1\}^{r_J}$ be the output of server $J$ to be broadcast, which is a function of $\xv_J$, and $Y_J$ be the corresponding random variable. For any set $P\subseteq N$, set $\yv_P \doteq (y_J,J\in P)$, and $\Yv_P \doteq (Y_J,J\in P)$. By convention, $\yv_{\emptyset}=\Yv_{\emptyset}=\emptyset$. 

For any DIC problem, we define 
a $(\uv,\rv) = ((u_i, i \in [n]), (r_J, J \in N))$ {\em distributed index code} by
\begin{itemize}
\item $2^n-1$ encoders, one for each server $J \in N$, such that $\phi_J: \prod_{j \in J} \{0,1\}^{u_j} \to \{0,1\}^{r_J}$ maps the messages $\xv_J$ in server $J$ to an $r_J$-bit sequence $y_J$, and
\item $n$ decoders, one for each receiver $i \in [n]$, such that $\psi_i: \prod_{J \in N} \{0,1\}^{r_J} \times \prod_{k \in A_i} \{0,1\}^{u_k} \to \{0,1\}^{u_i}$ maps the received sequences $\yv_N$ and the side information $\xv_{A_i}$ to $\hat{x}_i$.
\end{itemize}

We say that a rate--capacity tuple $(\mathbf{R}, \mathbf{C}) = ((R_i, i \in [n]),(C_J, J \in N))$ is achievable if for every $\epsilon > 0$, there exist a $(\uv,\rv)$ code and a positive integer $r$ such that %$\frac{u_i}{r}$, $i \in [n]$ and $\frac{r_J}{r}$, $J \in N$ satisfy
$R_i \le \frac{u_i}{r}$, $i \in [n]$, $C_J \geq \frac{r_J}{r}$, $J \in N,$
and that $\P\{ (\Xh_1,\ldots, \Xh_n) \ne (X_1, \ldots, X_n)\} \le \epsilon$. 

For a given link capacity tuple $\Cv$, the capacity region $\Cr(\Cv)$ is the closure of the set of all rate tuples $\Rv$ such that $(\Rv,\Cv)$ is achievable. The symmetric capacity is defined as $C_{\rm sym}(\Cv)=\max \{ R_{\rm sym}:(R_{\rm sym},\cdots,R_{\rm sym}) \in \Cr(\Cv) \}.$ The \emph{centralized index code}, the achievable rate tuple $\Rv$, the capacity region $\Cr$, and the symmetric capacity $C_{\rm sym}$ can be defined accordingly. 

Any CIC or DIC problem can be represented by a sequence $(i|j \in A_i)$, $i \in [n]$. For example, for $A_1 = \emptyset$, $A_2 = \{3\}$, and $A_3=\{2\}$, we write
$(1|-),\, (2|3),\, (3|2)$. 
It can also be represented by a side information graph $\Gcal$ with $n$ vertices, in which vertex $i$ represents message $i$, and a directed edge $(i,j)$ means that $i\in A_j$. 
For any nonempty message subset $S\subseteq [n]$,  $\Gcal|_S$ denotes the subgraph of $\Gcal$ induced by $S$. If $\Gcal|_{S}$ is acyclic, we simply say that the message group $S$ forms an acyclic structure or that $S$ is acyclic.

\section{Preliminaries}\label{sec:preliminaries}
\def\R{\mathcal R}
\newcommand{\ckj}{S_{K,J}}
\newcommand{\ck}{S_{K}}
\newcommand{\ckd}{S_{K}(\Dv)}
\newcommand{\proj}{\text{Proj}}
\newcommand{\CO}{\text{co}}
\newcommand{\D}{\mathcal{D}}
\newcommand{\ckjd}{S_{K,J}(\Dv)}
\newcommand{\ckjpdp}{S_{K,J}(P,\Dv)}
\newcommand{\rcc}{\Rr_\mathrm{CC}}
\newcommand{\rcce}{\Rr_\mathrm{CC}^{(e)}}
\newcommand{\bcc}{\b_\mathrm{CC}}
\newcommand{\bcce}{\b_\mathrm{CC}^{(e)}}

We briefly review the MAIS bound \cite{bar2011index} and the internal conflict bound \cite{maleki2014index,jafar2014topological}. %The former is stated below.
\begin{proposition}[MAIS bound, \cite{bar2011index}]         \label{prop:mais}
For the CIC problem $(i|A_i)$, $i\in [n]$, if $R_{\rm sym}$ is achievable, then 
\begin{align*}
R_{\rm sym}\le \min_{S\subseteq [n]:\text{$\Gcal|_{S}$ is acyclic}}\frac{1}{|S|}.
\end{align*}
\end{proposition}

For the internal conflict bound, we first re-state the definition of the alignment chain \cite{maleki2014index} in our notation as follows. 
\begin{definition}[Alignment Chain,\cite{maleki2014index}]         \label{def:ac}
For the CIC problem $(i|A_i)$, $i\in [n]$, messages $i(1),i(2),\cdots,i(m),i(m+1)$ and $k(1),k(2),\cdots,k(m)$ constitute an alignment chain of length $m$ denoted as
\begin{align}
\underline{i(1)} \xleftrightarrow{k(1)} i(2)
     \xleftrightarrow{k(2)} i(3) \cdots
     \xleftrightarrow{k(m)} \underline{i(m+1)},       \label{eq:ac}
\end{align}
if the conditions listed below are satisfied: 
\begin{enumerate}
\item $i(1)\in B_{i(m+1)}$ or $i(m+1)\in B_{i(1)}$;       \label{con:ac:1}
\item for any $j\in [m]$, we have $\{i(j),i(j+1) \} \subseteq B_{k(j)}$.        \label{con:ac:2}
\end{enumerate}
\end{definition}

For any alignment chain or any weighted alignment chain to be proposed later, we call the edge between $i(j)$ and $i(j+1)$ edge $j$. 
In Definition \ref{def:ac}, the two terminals $i(1)$ and $i(m+1)$ are underlined to indicate that $\{ i(1),i(m+1) \}$ is acyclic. 
Note that Definition \ref{def:ac} does not depend on the server setup, and thus also works for the DIC problem. 

For a CIC or DIC problem, if there exits at least one alignment chain, then we say that there is an \emph{internal conflict} between the two terminal messages $i(1)$ and $i(m+1)$ of the alignment chain and that the problem is \emph{internally conflicted}. 
The symmetric capacity for the CIC problems that are not internally conflicted is known \cite{arbabjolfaei2018fundamentals,maleki2014index,blasiak2013broadcasting}. For the internally conflicted CIC problems the following bound holds. 

\begin{proposition}[Internal conflict bound,\cite{maleki2014index,jafar2014topological}]\label{prop:ac}
For the internally conflicted CIC problem $(i|A_i)$, $i\in [n]$, if $R_{\rm sym}$ is achievable, then $R_{\rm sym}\le \frac{\Delta}{1+2\Delta}$,
where $\Delta$ denotes the minimum length of alignment chains for the problem.
\end{proposition}

In the rest of this paper, whenever we say a CIC or a DIC problem, we assume that the problem is internally conflicted.

\section{Main Results}  \label{sec:main:results}

\subsection{Improved Necessary Conditions for CIC}   \label{sec:gwac:cic}

We start by introducing a simple structure which will play a crucial role as the basic building block in the generalized alignment chains to be developed henceforth. 

\begin{definition}[Basic Tower]  \label{def:basic:tower}
For the CIC problem $(i|A_i)$, $i\in [n]$, messages $i(1),i(2) \in [n]$ and $k_1(1),\cdots,k_{h_1}(1) \in [n]$ constitute the following \emph{basic tower} $\Bcal_1$,
\begin{align*}
i(1) \xleftrightarrow{\substack{\quad k_{h_1}(1) \quad \\ \cdots \\ k_2(1) \\ k_1(1)}}_{\rm s} i(2),
\end{align*}
if $\{ i(1),i(2),k_1(1),\cdots,k_{\ell-1}(1) \} \subseteq B_{k_{\ell}(1)}$ for any $\ell \in [h_1]$.
\end{definition}

\begin{figure}[ht]
\begin{center}
\subfigure[][]{
\label{fig:basic:tower}
\includegraphics[scale=0.25]{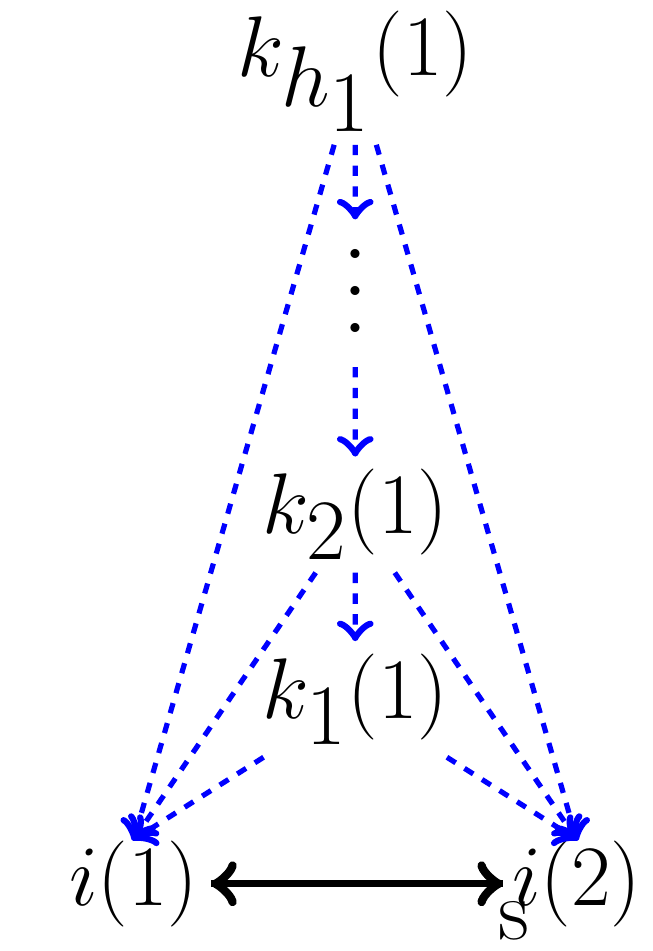}}
\subfigure[][]{
\label{fig:singleton:chain}
\includegraphics[scale=0.25]{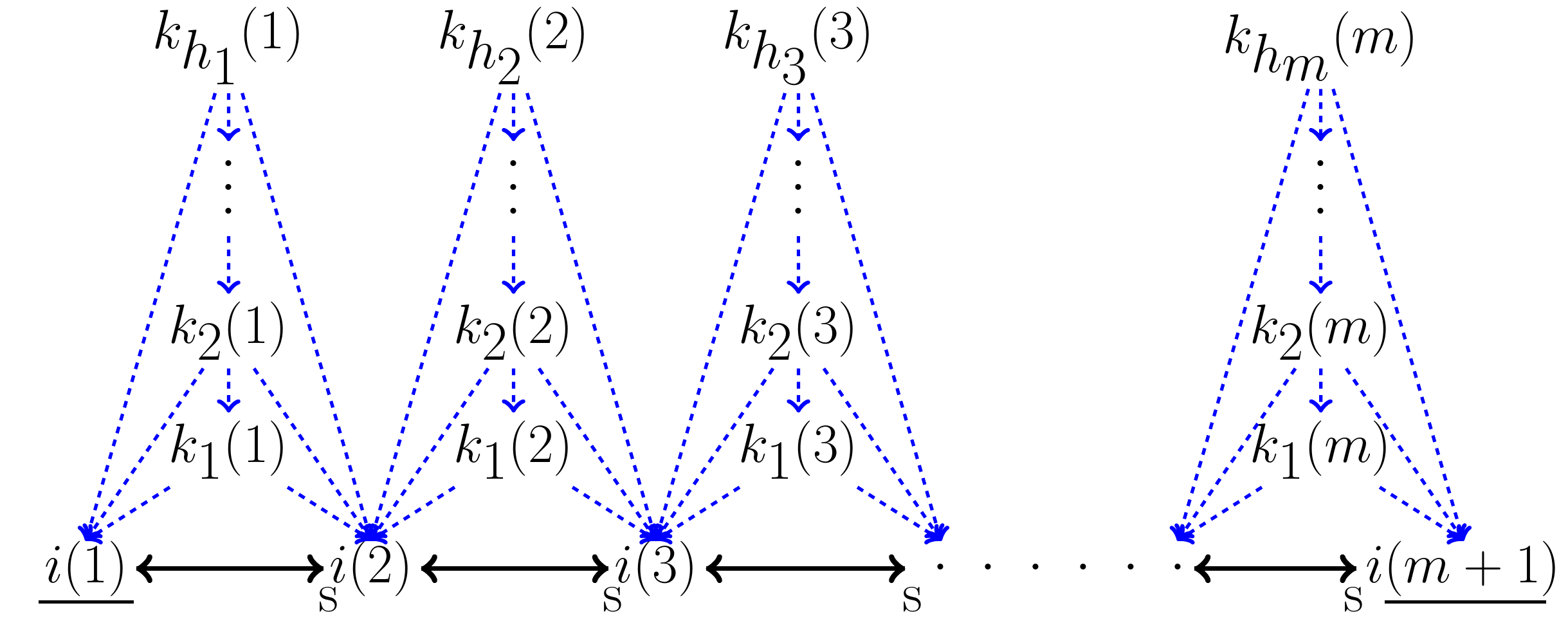}}
\caption{Schematic graphs for Definitions \ref{def:basic:tower} and \ref{def:wac:cic}: (a) a basic tower $\Bcal_1$, and (b) a singleton weighted alignment chain including $m$ concatenated basic towers. To help with understanding, we draw dashed arrows such that if there is a directed path of dashed arrows from message $a$ to $b$, then $b\in B_a$. For example in (a), $k_1(1) \in B_{k_{h_1}(1)}$ and there is a directed path as $k_{h_1}(1)\to k_{h_1-1}(1)\to \cdots \to k_{2}(1) \to k_{1}(1)$.}
\label{fig:1}
\end{center}
\end{figure}

A visualization of the above definition is given in Figure \ref{fig:basic:tower}. 
In the basic tower $\Bcal_1$, messages $i(1)$ and $i(2)$ are placed horizontally at the \emph{ground level} of the tower, and message $k_{\ell}(1)$ is placed on the $\ell$-th \emph{floor} for any $\ell \in [h_1]$, where $h_1$ is called the \emph{height} or the \emph{weight} of the tower. 
The receiver $k_{\ell}(1)$ who wants message $k_{\ell}(1)$ on the $\ell$-th floor cannot know any of the $k$-labeled messages on the lower floors nor the two $i$-labeled messages at the ground level as its side information. 
As a result, the message groups $\{i(1),k_1(1),\cdots,k_{h_1}(1)\}$ and $\{i(2),k_1(1),\cdots,k_{h_1}(1)\}$ are acyclic. 

%Now with the basic tower $\Bcal$ defined, we are ready to 
Now we propose the first generalization of the alignment chain, namely, the \emph{singleton weighted alignment chain}.%, which can be seen as a horizontal concatenation of the basic towers.

\begin{definition}[Singleton Weighted Alignment Chain] \label{def:wac:cic}
For the CIC problem $(i|A_i)$, $i\in [n]$, we have the following singleton weighted alignment chain, %\textcolor{blue}{I added $_{\rm s}$ to the arrows below and in Def. 2? should add them to figure 1 too?}
\begin{align*}
\underline{i(1)} \xleftrightarrow{\substack{k_{h_1}(1) \\ \cdots \\ k_{2}(1) \\ k_{1}(1)}}_{\rm s} i(2)
     \xleftrightarrow{\substack{k_{h_2}(2) \\ \cdots \\ k_{2}(2) \\ k_{1}(2)}}_{\rm s} \cdots 
     \xleftrightarrow{\substack{k_{h_m}(m) \\ \cdots \\ k_{2}(m) \\ k_{1}(m)}}_{\rm s} \underline{i(m+1)},      %     \label{eq:wac}
\end{align*}
denoted compactly as $i(1) \xleftrightarrow{\Ic,\Kcal }_{\rm s} i(m+1)$, where $\Ic \doteq \{ i(j):j\in [m+1] \}$, and $\Kcal \doteq \{ k_{\ell}(j):j\in [m],\ell\in [h_j] \}$, if the conditions listed below are satisfied: 
\begin{enumerate}
\item $i(1)\in B_{i(m+1)}$ or $i(m+1)\in B_{i(1)}$;       \label{con:wac:1}
\item for any $j\in [m]$, $\ell\in [h_j]$, $\{ i(j),i(j+1),k_1(j),\cdots,k_{\ell-1}(j) \} \subseteq B_{k_{\ell}(j)}$, i.e., messages $i(j),$ $i(j+1),k_{1}(j),\cdots,k_{h_j}(j)$ consitute a basic tower $\Bcal_{j}$.    \label{con:wac:2}
\end{enumerate}
\end{definition}

For simpler notation, in the rest of the paper, %for any singleton weighted alignment chain or any disjoint weighted alignment chain to be developed later, 
we use $K_{L}(j)$ to denote the message sequence $k_{\ell}(j)$, $\ell\in L$ for some $j\in [m]$, $L\subseteq [h_j]$, e.g. %$\{ K_{[h_j]}(j) \}=\{ k_{1}(j),k_{2}(j),\cdots,k_{h_j}(j) \}$. Hence 
the Condition \ref{con:wac:2} of Definition \ref{def:wac:cic} can be equivalently stated as that for any $j\in [m]$, $\ell \in [h_j]$, $\{ i(j),i(j+1),K_{[\ell-1]}(j) \} \subseteq B_{k_{\ell}(j)}$.

The singleton weighted alignment chain can be seen as a horizontal concatenation of $m$ basic towers, $\Bcal_1,\cdots,\Bcal_m$, such that the terminal message set of the chain, $\{ i(1),i(m+1) \}$, is acyclic. See Figure \ref{fig:singleton:chain} for visualization. 

We have the following theorem. 

\begin{theorem}\label{thm:wac:cic}
For the CIC problem $(i|A_i)$, $i\in [n]$, if $R_{\rm sym}$ is achievable, then for any of its singleton weighted alignment chains $i(1) \xleftrightarrow{\Ic,\Kcal }_{\rm s} i(m+1)$ we have
\begin{align}
R_{\rm sym}\leq \frac{m}{|\Ic|+|\Kcal|}=\frac{m}{1+m+\sum_{j\in [m]}h_j}.
\end{align}
\end{theorem}
The proof is presented in Appendix \ref{sec:proof:wac:cic}.

The original alignment chain can be viewed as a special case of the singleton weighted alignment chain where all the towers are of weight $1$. %with $h_j=1$ for any $j\in [m]$. 
Also, any acyclic message group constitutes a singleton weighted alignment chain with only one edge. %, or equivalently, as a special basic tower whose two horizontal terminals form an acyclic structure. 
Therefore, the internal conflict bound and the MAIS bound are strictly subsumed by Theorem \ref{thm:wac:cic}. 

Intuitively speaking, for every basic tower $\Bcal_j$ message sets $\{i(j),K_{[h_j]}(j)\}$ and $\{i(j+1),K_{[h_j]}(j)\}$ each form an acyclic structure, leading to a constraint on the symmetric rate, which can be captured by the MAIS bound. Yet, the singleton weighted alignment chain can further capture the concatenated relationship among these acyclic structures, which is not possible in the original MAIS bound.

Towards further generalization of the alignment chain, we introduce the following structure, which is defined based on the basic tower.

\begin{definition}[Crossing Tower]  \label{def:x:tower}
For the CIC problem $(i|A_i)$, $i\in [n]$, 
we have the following \emph{crossing tower},
\begin{align*}
i(1) \xleftrightarrow{\substack{k_{h_1}(1) \\ \cdots \\ k_{2}(1) \\ k_{1}(1)}}_{\rm s} \cdots i(j)
     \xleftrightarrow{\substack{k_{h_j}(j) \\ \cdots \\ k_{2}(j) \\ k_{1}(j)}}_{\rm c} i(j+1) \cdots
     \xleftrightarrow{\substack{k_{h_q}(q) \\ \cdots \\ k_{2}(q) \\ k_{1}(q)}}_{\rm s} i(q+1),      %     \label{eq:wac}
\end{align*}
denoted as $\Xcal_j$, if the conditions listed below are satisfied: 
\begin{enumerate}
\item for any $j'\in [q]\setminus \{j\}$, the message group $\{ i(j'),i(j'+1),K_{[h_{j'}]}(j') \}$ consitutes a basic tower $\Bcal_{j'}$;   \label{con:x:tower:1}
\item for any $\ell \in [h_j]$, $\{ K_{[\ell-1]}(j) \} \subseteq B_{k_{\ell}(j)}$;            \label{con:x:tower:2}
\item for any $\ell \in [h_j]$ there exist two integers $s_{\ell,j}\in [j]$, $t_{\ell,j}\in [j+1:q+1]$ such that  \label{con:x:tower:3}
    \begin{enumerate}
    \item $\{ i(s_{\ell,j}),i(t_{\ell,j}) \} \subseteq B_{k_{\ell}(j)}$;   \label{con:x:tower:31}
    \item for any $\ell_1< \ell_2 \in [h_j]$, we have $j = s_{1,j} \ge s_{\ell_1,j} \ge s_{\ell_2,j} \ge s_{h_j,j} = 1$, and $j+1 = t_{1,j} \le t_{\ell_1,j} \le t_{\ell_2,j} \le t_{h_j,j} = q+1$.   \label{con:x:tower:32}  
    \end{enumerate} 
\end{enumerate}
\end{definition}

For the crossing tower $\Xcal_j$ defined above, it has $q=q+1-1=t_{h_j,j}-s_{h_j,j}$ edges. We call edge $j$ the \emph{central} edge, and the message group $\{ i(j),i(j+1),K_{[h_j]}(j) \}$ the \emph{core}. 
Every other edge $j'\in [q]\setminus \{j\}$ corresponds to a basic tower $\Bcal_{j'}$. 
Note that we use different subscripts for the edges in the horizontal chain in the definition above to distinguish the two different types of edges.

Condition \ref{con:x:tower:3} in Definition \ref{def:x:tower} is described as follows. 
In the core, message $k_{\ell}(j)$ on the $\ell$-th floor has messages $i(s_{\ell,j})$ and $i(t_{\ell,j})$ to \emph{start} and \emph{terminate} its \emph{coverage}, respectively. In particular, for message $k_{1}(j)$ on the first floor, we have $i(s_{\ell,j})=i(j)$, and $i(t_{\ell,j})=i(j+1)$. 
The coverage of a message on a lower floor is within the range of the coverage of any message on a higher floor. 
We call the coverage of the message $k_{h_j}(j)$ on the top floor the \emph{total coverage} of the crossing tower, 
and define $G_{j}\doteq [s_{h_j,j}:t_{h_j,j}-1]$ denoting the set of edges located within the total coverage. 
Note that any basic tower $\Bcal_{j'}$ can be seen as a special crossing tower with $s_{\ell,j'}=j'$ and $t_{\ell,j'}=j'+1$ for any $\ell \in [h_{j'}]$, and hence $G_{j'}=\{j'\}$, and $|G_{j'}|=1$. 
Unless otherwise stated, when we say a crossing tower we assume that it is not a basic tower.

\iffalse
%Note that any basic tower can be seen as the shortest crossing tower with only the central edge and core, for which $j=t=1$, and $s_{j,\ell}=1$, $t_{j,\ell}=2$ for any $\ell \in [h_j]$. 
Also note that any basic tower $\Bcal_{j'}$ can be seen as a special crossing tower with only the central edge and the core, for which $s_{\ell}(j')=j'$, $t_{\ell}(j')=j'+1$ for any $\ell \in [h_{j'}]$. 
In the rest of the paper and unless otherwise stated, when we say a crossing tower we assume that it is not a basic tower. 

Condition \ref{con:x:tower:31} in Definition \ref{def:x:tower} is described as follows. Message $k_1(j)$ on the first floor of the central edge $j$ has message $i(j)$ to \emph{start} its \emph{coverage} and message $i(j+1)$ to \emph{terminate} its coverage. Message $k_{\ell}(j)$ on floor $\ell > 1$, has messages $i(s_{\ell}(j))$ and $i(t_{\ell}(j))$ to start and terminate its coverage, respectively. 
%Condition \ref{con:x:tower:32} ensures that the starting and terminating messages do not criss-cross. 
Condition \ref{con:x:tower:32} ensures that in core, the coverage of a message on a lower floor is within the range of the coverage of any other message on a higher floor. 
For any tower with central edge $j$, set $G_{j}\doteq [s_{h_j}(j):t_{h_j}(j)-1]$ denotes the \emph{total coverage} of the tower. Then for any basic tower $\Bcal_j$, $|G_j|=1$, and for any crossing tower $\Xcal_j$, $|G_j|\ge 2$. 
\fi

\begin{figure}[h]
\begin{center}
\includegraphics[scale=0.32]{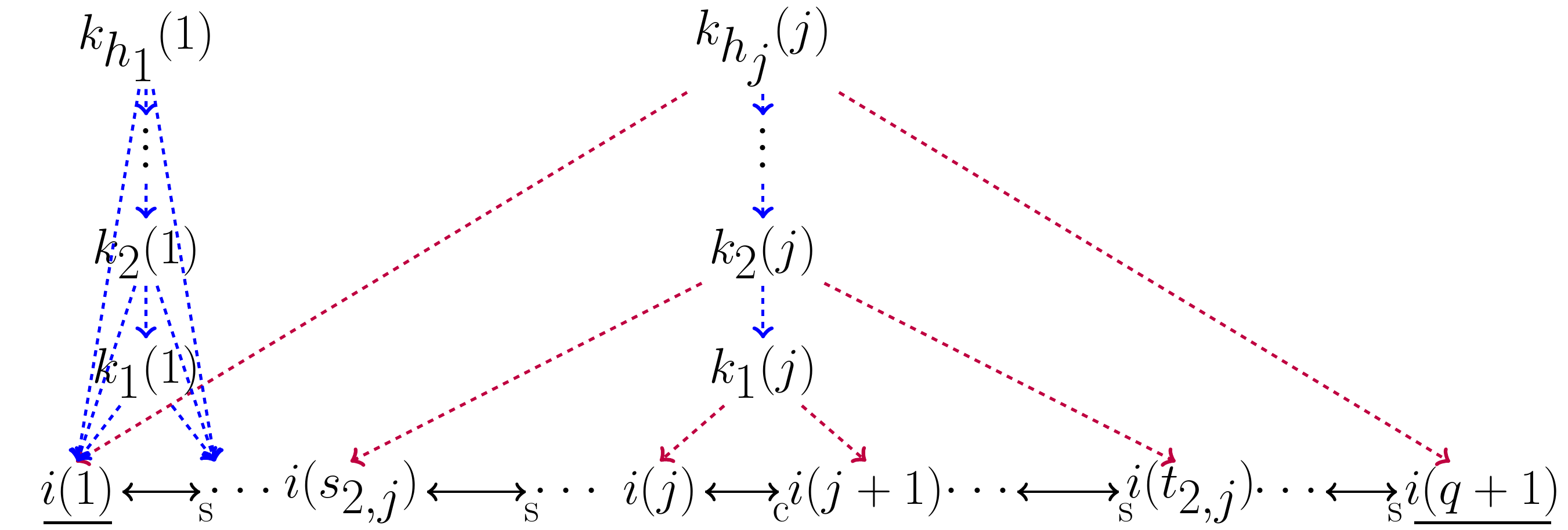}
\caption{A schematic graph for the crossing tower in Definition \ref{def:x:tower}. A directed path that contains arrows of only one color (either blue or purple) from message a to b indicates that $b\in B_a$. According to Condition \ref{con:x:tower:32} of Definition \ref{def:x:tower}, two purple arrows do not criss-cross.}
\label{fig:x:tower}
\end{center}
\end{figure}

For visualization of Definition \ref{def:x:tower}, see Figure \ref{fig:x:tower}. %Note that except for the central edge $j$ corresponding to the core, all other edges correspond to a basic tower. 
To avoid clutter, we only draw the leftmost basic tower of edge $1$ and the core of central edge $j$. Dashed arrows outgoing from the $k_{\ell}(j)$ messages in the core to their corresponding $i(s_{\ell,j})$ and $i(t_{\ell,j})$ messages in the horizontal chain are color-coded as purple, while all other dashed arrows are blue. 

%Now we are ready to present a further generalization of the singleton weighted alignment chain, namely, \emph{disjoint weighted alignment chain}. 

Now we are ready to present our most general alignment chain model, namely, \emph{disjoint weighted alignment chain}. 
\begin{definition}[Disjoint Weighted Alignment Chain] \label{def:gwac:cic}
For the CIC problem $(i|A_i)$, $i\in [n]$, we have the following disjoint weighted alignment chain, 
\begin{align*}
\underline{i(1)} \xleftrightarrow{\substack{k_{h_1}(1) \\ \cdots \\ k_{2}(1) \\ k_{1}(1)}} i(2)
     \xleftrightarrow{\substack{k_{h_2}(2) \\ \cdots \\ k_{2}(2) \\ k_{1}(2)}} i(3)
     \xleftrightarrow{\substack{k_{h_3}(3) \\ \cdots \\ k_{2}(3) \\ k_{1}(3)}} \cdots
     \xleftrightarrow{\substack{k_{h_m}(m) \\ \cdots \\ k_{2}(m) \\ k_{1}(m)}} \underline{i(m+1)},       \label{eq:gwac}
\end{align*}
denoted as $i(1)\xleftrightarrow{\Ic,\Kcal } i(m+1)$, if the conditions listed below are satisfied,
\begin{enumerate}
\item $i(1) \in B_{i(m+1)}$ or $i(m+1) \in B_{i(1)}$;       \label{con:gwac:1}
\item For every $j\in [m]$, message group $\{ i(j),i(j+1),K_{[h_j]}(j) \}$ constitutes either a basic tower $\Bcal_j$ or the core of a crossing tower $\Xcal_j$;   \label{con:gwac:2}
\item Set $M\doteq \{ j\in [m]:|G_j|\ge 2 \}$ denote the set of central edges of crossing towers, then for any $j_1 \neq j_2 \in M$, $G_{j_1}\cap G_{j_2}=\emptyset$, (i.e., the total coverage of different crossing towers must be disjoint).    \label{con:gwac:3}
%\item For each edge $j\in [m]$, it is a central edge of either a basic tower $\Bcal_j$ or a crossing tower $\Xcal_j$. Use $G_{j}\doteq [s_{j,h_j}:t_{j,h_j}-1]$ to denote the set of edges covered by a tower whose central edge is $j$. For basic tower $\Bcal_j$ $|G_j|=1$ and for crossing tower $\Xcal_j$ $|G_j|\ge 2$. Use $M\doteq \{ j\in [m]:|G_j|\ge 2 \}$ to denote the set of edges that is a central edge of a crossing tower. Then $\{ i(j),i(j+1),k_j(1),\cdots,k_j(\ell-1) \} \subseteq B_{k_j(\ell)}$ for any $j\in [m]\setminus M$, $\ell\in [h_j]$, and $j$ XXXXXXXXXXXXXXXXXXXX.
%\item For any $j_1 \neq j_2 \in M$, $G_{j_1}\cap G_{j_2}=\emptyset$ (i.e., the set of covered edges of different crossing towers must be disjoint).
%%
\end{enumerate}
\end{definition}

\begin{figure}[ht]
\begin{center}
\includegraphics[scale=0.31]{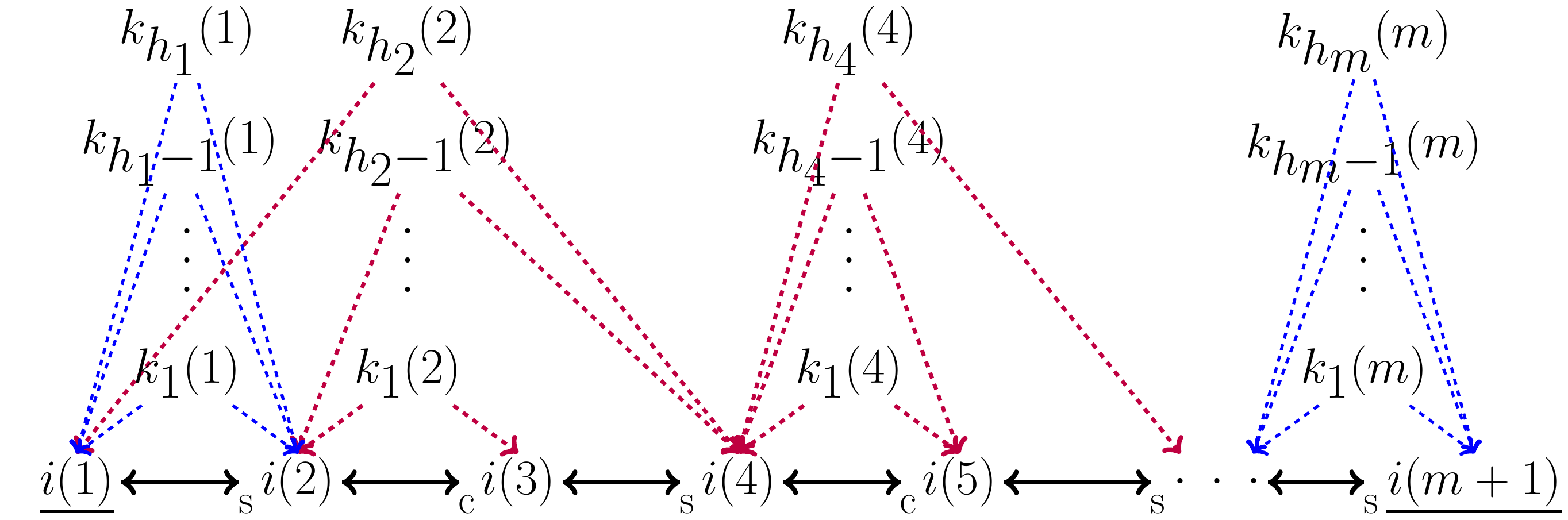}
\caption{A visualization example of the disjoint weighted alignment chain in Definition \ref{def:gwac:cic}. The arrows of the same color do not intersect.}
\label{fig:gwac:exm:general}%For any edge $j\in M=\{ j\in [m]:|G_j|\ge 2 \}$, the arrows outgoing from its associated $k_{j,\ell}$ messages are purple. For any edge $j\in [m]\setminus M$, the arrows outgoing from its associated $k_{j,\ell}$ messages are blue. 
\end{center}
\end{figure}

We remove any subscripts for the edges in the horizontal chain in Definition \ref{def:gwac:cic} since the positions of the basic and crossing towers are flexible in general. 
For a specific example of Definition \ref{def:gwac:cic}, see Figure \ref{fig:gwac:exm:general}. To help understanding, dashed arrows from $k_{\ell}(j)$ of some edges to the corresponding $i(s_{\ell,j})$ and $i(t_{\ell,j})$ are drawn. The dashed arrow is purple if the edge $j$ is in set $M$, %i.e., the edge is a central edge of a crossing tower, 
and blue otherwise. 
Definitions \ref{def:x:tower} and \ref{def:gwac:cic} jointly ensure that two purple dashed arrows can never criss-cross. 
%Condition \ref{con:x:tower:32} in Definition \ref{def:x:tower} and Condition \ref{con:gwac:3} in Definition \ref{def:gwac:cic} jointly ensure that two purple dashed arrows can never criss-cross. 

%Note that the basic towers $\Bc_{j'}, j'\in [m]\setminus M$ in a disjoint weighted alignment chain are either located within the total coverage $G_j$ of a crossing tower $\Xcal_j$ (e.g., the leftmost $\Bc_1$ in Figure \ref{fig:gwac:exm:general}) or located outside the total coverage of any crossing towers (e.g., the rightmost $\Bc_m$ in Figure \ref{fig:gwac:exm:general}). 
Define $M'\doteq [m]\setminus (\bigcup_{j\in M}G_j)$ as the set of edges located outside the total coverage of any crossing tower. 
Then the disjoint weighted alignment chain can be seen as a horizontal concatenation of the crossing towers $\Xcal_j$, $j\in M$ and the basic towers $\Bc_{j'}$, $j'\in M'$, such that $\{ i(1),i(m+1) \}$ is acyclic. 
Clearly, any singleton weighted alignment chain can be seen as a special disjoint weighted alignment chain, for which $M=\emptyset$, and $M'=[m]$.

%\vspace{-5mm}

We have the following theorem. 
%%%%%%%%%%%%%%%%%%
\begin{theorem}\label{thm:gwac:cic}
For the CIC problem $(i|A_i)$, $i\in [n]$, if $R_{\rm sym}$ is achievable, then for any of its disjoint weighted alignment chains $i(1)\xleftrightarrow{\Ic,\Kcal } i(m+1)$ we have
\begin{align}
R_{\rm sym}\leq \frac{m}{|\Ic|+|\Kcal|}=\frac{m}{1+m+\sum_{j\in [m]}h_j}.
\end{align}
\end{theorem}

The proof is presented in Appendix \ref{sec:modular:gwac:cic}.

%%%%%%%%%%%%%%%%%%%%%%%%%%%%%%%%%%%%%%%%%%%%%
%%%%%%%%%%%%%%%%%%%%%%%%%%%%%%%%%%%%%%%%%%%%%

For the CIC problem $(i|A_i)$, $i\in [n]$, let $R_{\rm DW}$, $R_{\rm SW}$, $R_{\Delta}$, and $R_{\rm MAIS}$ denote the upper bounds given by Theorem \ref{thm:gwac:cic}, Theorem \ref{thm:wac:cic}, Proposition \ref{prop:ac}, and Proposition \ref{prop:mais}, respectively. 
In the following proposition, we formalize the relationships between these upper bounds, which were alluded to earlier in this subsection. 

\begin{proposition}   \label{prop:hierarchy:cic}
$R_{\rm DW}\le R_{\rm SW}\le R_{\Delta}$, and $R_{\rm SW}\le R_{\rm MAIS}$.
\end{proposition}

\begin{IEEEproof}
Any alignment chain can be seen as a singleton weighted alignment chain and any singleton weighted alignment chain can be seen as a disjoint weighted alignment chain. Therefore, it is clear that $R_{\rm DW}\le R_{\rm SW}\le R_{\Delta}$. 

Set $s=1/R_{\rm MAIS}$, then there exists an acyclic message set of size $s$, say $\{ i(1),i(2),\cdots,i(s) \}$. In other words, 
\begin{align}
\{i(1),\cdots,i(\ell-1)\} \subseteq B_{i(\ell)}, \qquad \forall \ell \in [s].
\end{align}
Therefore, we have the following one-edge singleton weighted alignment chain,
\begin{align}
\underline{i(1)} \xleftrightarrow{\substack{\quad i(s) \quad \\ \cdots \\ i(3)}} _{\rm s}\underline{i(2)},
\end{align}
and thus by Theorem \ref{thm:wac:cic}, we have $R_{\rm SW}\le \frac{1}{s}=R_{\rm MAIS}$.
\end{IEEEproof}

The relationships in Proposition \ref{prop:hierarchy:cic} can sometimes be strict.
\begin{example}
Consider the $6$-message CIC problem
\begin{align*}
\begin{array}{ccc}
(1|2,3,4,6), &(2|4,5,6), &(3|1,2,4,5,6) \\
(4|1,2,6), &(5|2,3,4,6), &(6|-).
\end{array}
\end{align*}
For this problem, $R_{\Delta}=R_{\rm MAIS}=\frac{1}{3}$. 
However, we have the following singleton weighted alignment chain,
\begin{align*}
\underline{1} \xleftrightarrow{\substack{\quad 6 \quad \\ 2}}_{\rm s} 3 \xleftrightarrow{\substack{\quad 6 \quad \\ 4}}_{\rm s} \underline{5},
\end{align*}
and thus by Theorem \ref{thm:wac:cic} we have $R_{\rm sym}\le \frac{2}{3+4}=\frac{2}{7}$, which matches the composite coding lower bound \cite{Arbabjolfaei--Bandemer--Kim--Sasoglu--Wang2013} on the symmetric capacity. Therefore, for this problem, we have $C_{\rm sym}=R_{\rm DW}=R_{\rm SW}=\frac{2}{7}<R_{\Delta}=R_{\rm MAIS}=\frac{1}{3}.$ Note that in the above chain, message $6$ appears twice in two different towers, which is allowed. 
\end{example}

\begin{example}
Consider the $10$-message CIC problem 
\begin{align*}
\begin{array}{cc}
(1|3,4,5,6,7,8,9,10), & (2|3,4,5,6,7,8,9,10)  \\
(3|1,2,4,5,6,7,8,9,10), & (4|1,2,3,5,6,7,8,9,10), \\
(5|1,3,6,7,8,9,10),  & (6|2,4,5,7,8,9,10), \\
(7|1,2,5,6,8,9,10),  & (8|1,3,6,7,9,10), \\
(9|2,3,5,7,8,10), & (10|1,2,5,6,8,9).
\end{array}
\end{align*}
For this problem, $R_{\Delta}=R_{\rm MAIS}=R_{\rm SW}=\frac{1}{3}$. However, we have the following disjoint weighted alignment chain,
\begin{align*}
\underline{1} \xleftrightarrow{\substack{\quad 9 \quad \\ 6}}_{\rm c} 3 \xleftrightarrow{\substack{\quad 10 \quad \\ 7}}_{\rm s} 4 \xleftrightarrow{\substack{\quad 8 \quad \\ 5}}_{\rm s} \underline{2},   %\label{eq:gwac:cic:10message}
\end{align*}
and thus by Theorem \ref{thm:gwac:cic} we have $R_{\rm sym}\le \frac{3}{4+6}=\frac{3}{10}$, which matches the composite coding lower bound \cite{Arbabjolfaei--Bandemer--Kim--Sasoglu--Wang2013} on the symmetric capacity. Therefore, for this problem, we have 
$$C_{\rm sym}=R_{\rm DW}=\frac{3}{10}<R_{\rm SW}=R_{\Delta}=R_{\rm MAIS}=\frac{1}{3}.$$ 
Note that the above disjoint weighted alignment chain is not a singleton weighted alignment chain due to the existence of the crossing tower $\Xcal_1$ whose central edge is edge $1$, i.e., the edge between messages $1$ and $3$. For $\Xcal_1$, message $6$ on the first floor of the core has messages $1$ and $3$ to start and terminate its coverage, respectively, and message $9$ on the second floor has messages $1$ and $4$ to start and terminate its coverage, respectively, and thus $|G_1|=2$. 
%$B_{9}=\{1,4,6\} \not \supseteq \{ 1,3,6 \}$.
%Note that \eqref{eq:gwac:cic:10message} is not a singleton weighted alignment chain according to Definition \ref{def:wac:cic} as $B_{9}=\{1,4,6\} \not \supseteq \{ 1,3,6 \}$.
\end{example}

The new bounds can be also be useful for solving some large problems, for which the more general PM bound is computationally infeasible in practice. 
\begin{example}    \label{exm:gwac:cic:n17}
Consider the $17$-message CIC problem as follows, denoted by $(i|B_i)$, $i\in [n]$ rather than $(i|A_i)$, $i\in [n]$ for brevity,
\begin{align*}
&(1|6), (2|7,8), (3|8,11,17), (4|-), (5|-), (6|1), (7|1,2),  \\
&(8|1,2,3,4,7), (9|2,3), (10|1,4,9), (11|3,4,8), (12|5,6),  \\
&(13|4,5), (14|4,6,7,13,15,17), (15|-), (16|5,6,12), (17|8).
\end{align*}
For this problem, $R_{\Delta}=R_{\rm MAIS}=R_{\rm SW}=\frac{1}{3}$. However, we have the following disjoint weighted alignment chain, 
\begin{align*}
\underline{1} \xleftrightarrow{\substack{\enskip 8 \enskip \\ 7}}_{\rm s} 2 \xleftrightarrow{\substack{\enskip 10 \enskip \\ 9}}_{\rm c} 3 \xleftrightarrow{\substack{\enskip 11 \enskip \\ 8}}_{\rm s} 4 \xleftrightarrow{\substack{\enskip 14 \enskip \\ 13}}_{\rm c} 5 \xleftrightarrow{\substack{\enskip 16 \enskip \\ 12}}_{\rm s} \underline{6},  
\end{align*}
and thus by Theorem \ref{thm:gwac:cic} we have $R_{\rm sym}\le R_{\rm DW}\le \frac{5}{6+10}=\frac{5}{16}<\frac{1}{3}$, which matches the composite coding lower bound \cite{Arbabjolfaei--Bandemer--Kim--Sasoglu--Wang2013} on $C_{\rm sym}$. Note that the above chain is not a singleton weighted alignment chain due to the existences of the two crossing towers $\Xcal_2$ and $\Xcal_4$, whose central edges are edge $2$ that is between messages $2$ and $3$, and edge $4$ that is between messages $4$ and $5$, respectively. 
For $\Xcal_2$, its total coverage starts at message $1$ and terminates at message $4$, and thus $|G_2|=3$. For $\Xcal_4$, its total coverage starts at message $4$ and terminates at message $6$, and thus $|G_4|=2$. 
%For example, $B_{10}=\{ 1,4,9 \}\not \supseteq \{ 2,3,9 \}$.  % as some $p_{j,\ell}\neq j-1$, e.g, .
%\textcolor{red}{can you specify one red dashed edge going across?}
\end{example}

%%%%%%%%%%%%%%%%%%%%%%%%%%%%%%%%%%%%%%%%%%%%%%%%%%%%%%%
%%%%%%%%%%%%%%%%%%%%%%%%%%%%%%%%%%%%%%%%%%%%%%%%%%%%%%%
%%%%%%%%%%%%%%%%%%%%%%%%%%%%%%%%%%%%%%%%%%%%%%%%%%%%%%%
%%%%%%%%%%%%%%%%%%%%%%%%%%%%%%%%%%%%%%%%%%%%%%%%%%%%%%%
%%%%%%%%%%%%%%%%%%%%%%%%%%%%%%%%%%%%%%%%%%%%%%%%%%%%%%%

%\vspace{2mm}
%\newpage
\subsection{Improved Necessary Conditions for DIC}       \label{sec:gwac:dic}

Definitions \ref{def:ac}-\ref{def:gwac:cic} also apply to the DIC problem. In the following we extend Theorems \ref{thm:wac:cic} and \ref{thm:gwac:cic} to Theorems \ref{thm:wac:dic} and \ref{thm:gwac:dic}, respectively. Recall that $G_{j}=[s_{h_j,j}:t_{h_j,j}-1]$, $M=\{ j\in [m]:|G_j|\ge 2 \}$, and $M'=[m]\setminus (\bigcup_{j\in M}G_j)$.

\begin{theorem}\label{thm:wac:dic}
For the DIC problem $(i|A_i)$, $i\in [n]$ with link capacity tuple $\Cv$, if $R_{\rm sym}$ is achievable, then for any of its singleton weighted alignment chains $i(1)\xleftrightarrow{\Ic,\Kcal }_{\rm s} i(m+1)$ we have
\begin{align*}
R_{\rm sym}\leq &\frac{1}{|\Ic|+|\Kcal|}\sum_{j\in [m]}\sum_{\substack{J\in N: J\cap \{ i(j),i(j+1),K_{[h_j]}(j)\} \neq \emptyset, \\ J\cap \{ i(j),i(m+1),K_{[h_j]}(j)\} \neq \emptyset}}C_J.   %\label{eq:wac:dic}
\end{align*}
\end{theorem}

\begin{theorem}\label{thm:gwac:dic}
For the DIC problem $(i|A_i)$, $i\in [n]$ with link capacity tuple $\Cv$, if $R_{\rm sym}$ is achievable, then for any of its disjoint weighted alignment chains $i(1)\xleftrightarrow{\Ic,\Kcal } i(m+1)$ we have
\begin{align}
R_{\rm sym}   
\le &\frac{1}{|\Ic|+|\Kcal|}\big(
\sum_{\substack{j\in M\cup M'}}\sum_{\substack{J\in N: J\cap T_1(j)\neq \emptyset, J\cap T_2(j)\neq \emptyset}}C_J   \nonumber   \\
&+\sum_{j\in M}\sum_{\ell \in [2:h_j]}( \sum_{j' \in [s_{\ell,j}:s_{\ell-1,j}-1]}\sum_{\substack{J\in N: J\cap T_3(j')\neq \emptyset,\\ J\cap T_4(j,\ell,j')\neq \emptyset}}C_J       \nonumber  \\ 
&+\sum_{j' \in [t_{\ell-1,j}:t_{\ell,j}-1]}\sum_{\substack{J\in N: J\cap T_3(j')\neq \emptyset,\\ J\cap T_5(j,\ell,j')\neq \emptyset}}C_J ) \big),  \label{eq:gwac:dic}
\end{align}
where 
\begin{align*}
&T_1(j)=\{ i(s_{h_j,j}),i(t_{h_j,j}),K_{[h_j]}(j) \},   \nonumber    \\
&T_2(j)=\{ i(s_{h_j,j}),i(m+1),K_{[h_j]}(j) \},      \nonumber     \\
&T_3(j')=\{ i(j'),i(j'+1),K_{[h_{j'}]}(j') \},   \nonumber   \\
&T_4(j,\ell,j')=\{ i(j'),i(s_{\ell-1,j}),K_{[h_{j'}]}(j') \},     \nonumber    \\
&T_5(j,\ell,j')=\{ i(j'),i(t_{\ell,j}),K_{[h_{j'}]}(j') \}.
\end{align*}
\iffalse
\begin{align}
&R_{\rm sym}\leq 
\frac{1}{|\Ic|+|\Kcal|}\big(
\sum_{j\in M}\sum_{\ell \in [2:h_j]}( \sum_{j' \in [s_{\ell,j}:s_{\ell-1,j}-1]}\sum_{\substack{J\cap T_3\neq \emptyset,\\ J\cap T_4\neq \emptyset}}C_J       \nonumber  \\ 
&+\sum_{j' \in [t_{\ell-1,j}:t_{\ell,j}-1]}\sum_{\substack{J\cap T_3\neq \emptyset,\\ J\cap T_5\neq \emptyset}}C_J )+\sum_{\substack{j\in \\M\cup M'}}\sum_{\substack{J\cap T_1\neq \emptyset,\\ J\cap T_2\neq \emptyset}}C_J  \big),  \label{eq:gwac:dic}
\end{align}
where 
\begin{align*}
T_1&=\{ i(s_{h_j,j}),i(t_{h_j,j}),K_{[h_j]}(j) \},   \nonumber    \\
T_2&=\{ i(s_{h_j,j}),i(m+1),K_{[h_j]}(j) \},      \nonumber     \\
T_3&=\{ i(j'),i(j'+1),K_{[h_{j'}]}(j') \},   \nonumber   \\
T_4&=\{ i(j'),i(s_{\ell-1,j}),K_{[h_{j'}]}(j') \},     \nonumber    \\
T_5&=\{ i(j'),i(t_{\ell,j}),K_{[h_{j'}]}(j') \}.
\end{align*}
\fi
\end{theorem}

The proofs for Theorems \ref{thm:wac:dic} and \ref{thm:gwac:dic} are presented in Appendices \ref{sec:proof:wac:dic} and \ref{sec:proof:gwac:dic}, respectively. Note that these proofs are similar in principle, but more involved compared to their centralized counterparts. 
%which are similar but more complicated compared to the proofs of Theorems \ref{thm:wac:cic} and \ref{thm:gwac:cic}, respectively. 
%is presented in Section \ref{sec:proof:gwac:dic}. 
%The proofs for Theorems \ref{thm:wac:dic} and \ref{thm:gwac:dic} are presented in 

\begin{example}
Consider the following DIC problem with $n=5$ messages and equal link capacities $C_J=1,J\in N$,
\begin{align*}
(1|2,3,4,5), (2|1,3,4,5), (3|2,4,5), (4|3,5), (5|1,4).
\end{align*}
For this problem, 
there exists a singleton weighted alignment chain as $\underline{1} \xleftrightarrow{\quad 4 \quad }_{\rm s} 2 \xleftrightarrow{\quad 5 \quad}_{\rm s} \underline{3}$, 
and thus by Theorem \ref{thm:wac:dic}, 
\begin{align*}
R_{\rm sym}\le \frac{1}{5}(\sum_{\substack{J\in N: J\cap \{ 1,2,4 \}\neq \emptyset,\\ J\cap \{ 1,3,4 \}\neq \emptyset}}1+\sum_{\substack{J\in N:  J\cap \{ 2,3,5 \}\neq \emptyset}}1)=\frac{54}{5},
\end{align*}
%which matches the composite coding lower bound \cite{Arbabjolfaei--Bandemer--Kim--Sasoglu--Wang2013} on the symmetric capacity. 
which matches the distributed composite coding lower bound \cite{liu2018capacity} on the symmetric capacity. 
\end{example}

%%%%%%%%%%%%%%%%%%%%%%%%%%%%%%%%%%%%%%%%%%%%%%%%%%%%%%%%%%%%%%%%%%%%%%%%%%%%%%%%%%%%%%%%%%%%%%%%%%%%%%%%%%%%%%%%%%%%%%%%%%%%%%%%%%%%%%%%
%\vspace{-1mm}
\subsection{A $9$-Message CIC Problem Which Needs Non-Shannon-Type Inequalities for Capacity Characterization}    \label{sec:nonshannon}
%Here we present a separate, but related result on the converse for the CIC problem, which can be interesting on its own. 
We present a $9$-message CIC problem, which is the CIC problem with the smallest number of messages $n$ identified so far, for which non-Shannon-type inequalities are necessary to derive a tighter converse on the capacity region. 
\begin{example}   \label{exm:zy}
Consider the following $9$-message CIC problem, denoted by $(i|B_i)$, $i\in [n]$,
\begin{align*}
&(1|2), (2|1,5,8), (3|-), (4|-), (5|2,4,8),  \\
&(6|1,3), (7|3,4), (8|2,3,5), (9|1,4,6).
\end{align*}
%The tightest upper bound on the achievable sum-rate givable by Shannon-type inequalities is $\sum_{i\in [n]}R_i\le 19/6$, which is computed using the PM bound.
The PM bound\footnote{The PM bound, including all decoding constraints, is as tight as the bound utilizing all Shannon-type inequalities. See the full arXiv version of \cite{liu:vellambi:kim:sadeghi:itw18}.} gives $\sum_{i\in [n]}R_i\le 19/6$. 
However, applying Zhang-Yeung non-Shannon-type inequalities \cite{zhang1998characterization} to the problem, the upper bound can be further tightened to
\begin{align}
\sum_{i\in [n]}R_i\leq \frac{25}{8}.  \label{eq:zy:n9:goal}
\end{align}
Note that for this problem we are focusing on the achievable sum-rate, rather than the symmetric rate for which Shannon-type inequalities are capable of giving tight upper bound. 
The proof of \eqref{eq:zy:n9:goal} is presented in Appendix \ref{sec:proof:zy}. 
\end{example}
\appendices
%\section{Proofs}   \label{sec:proofs}

\section{Proof of Theorem \ref{thm:wac:cic}} \label{sec:proof:wac:cic}
%\section{Defining set functions to be used in the proofs}

Before presenting the proofs for Theorems \ref{thm:wac:cic}-\ref{thm:gwac:dic}, and \eqref{eq:zy:n9:goal}, we first introduce two set functions for the CIC and the DIC problems, which will be extensively used later. 

For the CIC problem $(i|A_i)$, $i\in [n]$, define set function $g(S)\doteq \frac{1}{r}H(Y_{[n]}|\Xv_{S^c})$, for any $S\subseteq [n]$. Clearly, $g(\emptyset)=0$.  It can be verified that for any $S,S'\subseteq [n]$, the set function $g(S)$ has the following two properties, %is monotonically non-decreasing, $g(S) \leq g(S')$ for $S \subseteq S'\subseteq [n]$, and submodular, $g(S\cap S') + g(S\cup S') \leq g(S) + g(S')$, $\forall S,S'\subseteq [n]$. 
\begin{align*}
&g(S)\le g(S'),  \qquad \text{if $S\subseteq S'$,}    \\
&g(S\cap S') + g(S\cup S') \leq g(S) + g(S'),
\end{align*}
which we refer to as the monotonicity and submodularity of $g(S)$, respectively. 
Also, $g(S)\le 1,\forall S\subseteq [n]$. 
Note that set functions equivalent to $g(S)$ have been used in \cite{Arbabjolfaei--Bandemer--Kim--Sasoglu--Wang2013,liu:vellambi:kim:sadeghi:itw18} to study the PM bound for the CIC problem. In particular, detailed proofs for the properties of $g(S)$ can be found in \cite[Section 5.2]{arbabjolfaei2018fundamentals}.

Similarly, for the DIC problem $(i|A_i)$, $i\in [n]$ with link capacity tuple $\Cv$, define set function $f(L;S)\doteq \frac{1}{r}H(\Yv_{\{ J\in N:J\cap L\neq \emptyset \}}|\Xv_{S^c})$ for any $L,S\subseteq [n]$. Obviously, $f(\emptyset;S)=f(L;\emptyset)=0$. It can be verified that for any $L,L',S,S'\subseteq [n]$, the two-argument set function $f(L;S)$ has the following two properties,
\begin{align*}
%&f(\emptyset,S)=f(L,\emptyset)=0,     \\
&f(L;S)\le f(L';S'), \qquad \text{if $L\subseteq L'$, $S\subseteq S'$,}  \\
&f(L\cap L';S\cup S')+f(L\cup L';S\cap S')\le f(L;S)+f(L';S'),
\end{align*}
which we refer to as the monotonicity and submodularity of $f(L;S)$, respectively. Also, for any $L,S\subseteq [n]$, we have
\begin{align}
&f([n];S)=f(S;S), \label{eq:f:linking}  \\
&f(L;S)\le \sum_{J\in N: J\cap L\neq \emptyset,J\cap S\neq \emptyset}C_J.   \label{eq:f:capacity:constraint}
\end{align} 
Note that a two-argument set function equivalent to $f(L;S)$ defined above has been introduced in \cite[Section V]{isit:2017}, where the proofs for its properties are also presented. 

We sometimes use $g(i,i\in S)$ to denote $g(S)$. Similarly, $f(L;S)$ can be written in equivalent forms as $f(L;i,i\in S)$, $f(k,k\in L;S)$, and $f(k,k\in L;i,i\in S)$. 

For the two set functions $g(S)$ and $f(L;S)$, the following two lemmas hold. 
%The lemma below holds universally for any CIC problem. 

\begin{lemma}[\cite{liu:vellambi:kim:sadeghi:itw18}] \label{lem:decoding:Bi}
Consider the CIC problem $(i|A_i)$, $i\in [n]$. For any $i\in [n]$, we have
\begin{align}
R_i+g(B)=g(B\cup \{i\}), \qquad \forall B\subseteq B_i.
\end{align}
Particularly, when $B=\emptyset$, $g(B)=0$, and thus $R_i=g(\{ i \})$.
\end{lemma}

\begin{lemma}  \label{lem:decoding:Bi:dic}
Consider the DIC problem $(i|A_i)$, $i\in [n]$ with link capacity tuple $\Cv$. For any $i\in [n]$, we have
\begin{align}
R_i+f([n];B)=f([n];B\cup \{i\}), \qquad \forall B\subseteq B_i.
\end{align}
Particularly, as $f([n];\emptyset)=0$, we have $R_i=f([n];\{ i \})$.
\end{lemma}

Lemma \ref{lem:decoding:Bi:dic} can be proved similarly to Lemma \ref{lem:decoding:Bi}.

%%%%%%%%%%%%%%%%%%%%%%%%%%%%%%%%%%%%%%%%%%%%
%%%%%%%%%%%%%%%%%%%%%%%%%%%%%%%%%%%%%%%%%%%%
%%%%%%%%%%%%%%%%%%%%%%%%%%%%%%%%%%%%%%%%%%%%
%THEOREM 1

%\newpage
%\section{Proof of Theorem \ref{thm:wac:cic}} \label{sec:proof:wac:cic}
%b_{j}(1),\cdots,b_{j}(h_j)

%We first present the following lemmas. % for any basic tower. 
To show Theorem \ref{thm:wac:cic}, we further introduce the following lemmas. 
\begin{lemma}   \label{lem:basic:tower}
For the CIC problem $(i|A_i)$, $i\in [n]$ with the basic tower $\Bc_j$ constituted by the message group $\{ i(j),i(j+1),K_{[h_j]}(j) \}$, 
we have
\begin{align}
g(i(j),i(j+1))+\sum_{\ell \in [h_j]}R_{k_{\ell}(j)}\le 1.   \nonumber  %\label{eq:lem:basic:tower}
\end{align}
\end{lemma}
\begin{IEEEproof}
For the basic tower $\Bc_j$, we have
\begin{align}
1&\ge g(i(j),i(j+1),K_{[h_j]}(j))    \nonumber   \\
  &=g(i(j),i(j+1),K_{[h_j-1]}(j))+R_{k_{h_j}(j)}     \nonumber   \\
  &=g(i(j),i(j+1),K_{[h_j-2]}(j))+R_{k_{h_j-1}(j)}+R_{k_{h_j}(j)}     \nonumber   \\
  &\cdots    \nonumber   \\
  &=g(i(j),i(j+1))+\sum_{\ell \in [h_j]}R_{k_{\ell}(j)},   \nonumber
\end{align}
where the inequality is due to the fact that $g(S)\le 1$, $\forall S\subseteq [n]$, and the equalities follow from Lemma \ref{lem:decoding:Bi} with $\{ i(j),i(j+1),K_{[\ell-1]}(j) \} \subseteq B_{k_{\ell}(j)}$, $\forall \ell \in [h_j]$ by Definition \ref{def:basic:tower}.% the Condition \ref{con:wac:2} of Definition \ref{def:wac:cic}. 
\end{IEEEproof}

\begin{lemma}   \label{lem:horizontal}
For the CIC problem $(i|A_i)$, $i\in [n]$ with the weighted alignment chain $i(1)\xleftrightarrow{\Ic,\Kcal } i(m+1)$, for any two integers $a<b\in [m]$, we have
\begin{align}
\sum_{j\in [a:b]}g(i(j),i(j+1))\ge \sum_{j\in [a+1:b]}R_{i(j)}+g(i(a),i(b+1)).   \nonumber   
%&\ge \sum_{j\in [a+1:b]}R_{i(j)}+g(i(a),i(a+1),\cdots,i(b+1)).
\end{align}
\end{lemma}
\begin{IEEEproof}
We have 
\begin{align}
&\sum_{j\in [a:b]}g(i(j),i(j+1))   \nonumber   \\
&=g(i(a),i(a+1))+g(i(a+1),i(a+2)) \nonumber   \\   
&\quad +g(i(a+2),i(a+3))+\cdots+g(i(b),i(b+1))   \nonumber    \\
&\ge R_{i(a+1)}+g(i(a),i(a+1),i(a+2))   \nonumber   \\
&\quad +g(i(a+2),i(a+3))+\cdots+g(i(b),i(b+1))   \nonumber    \\
&\ge R_{i(a+1)}+R_{i(a+2)}+g(i(a),i(a+1),i(a+2),i(a+3))   \nonumber   \\
&\quad +\cdots+g(i(b),i(b+1))   \nonumber    \\
&\cdots    \nonumber    \\
&\ge R_{i(a+1)}+R_{i(a+2)}+\cdots+R_{i(b)}   \nonumber   \\
&\quad +g(i(a),i(a+1),i(a+2),\cdots,i(b),i(b+1))   \nonumber   \\
&\ge \sum_{j\in [a+1:b]}R_{i(j)}+g(i(a),i(b+1)), 
\end{align}
%where we show the derivation of the first inequality explicitly as follows:
where the first inequality follows from the following derivation:
\begin{align*}
&g(i(a),i(a+1)) + g(i(a+1), i(a+2))    \\
&\ge g(i(a+1)) + g(i(a), i(a+1), i(a+2))     \\
&=R_{i(a+1)}+ g(i(a), i(a+1), i(a+2)),
\end{align*}
where the inequality is due to the submodularity of $g(S)$ and the equality is due to Lemma 1. The remaining inequalities, except for the last one follow similarly. The last inequality above is due to the monotonicity of $g(S)$. 
\end{IEEEproof}

Now we are ready to present the proof for Theorem \ref{thm:wac:cic}. 
\begin{IEEEproof}
Consider the singleton weighted alignment chain $i(1)\xleftrightarrow{\Ic,\Kc}_{\rm s} i(m+1)$ for a CIC problem, which can be seen as a concatenation of $m$ basic towers $\Bc_1,\Bc_2,\cdots,\Bc_m$. %, where $\Bc_j$ is formed by the message group $\{ i(j),i(j+1),k_j(1),\cdots,k_j(h_j) \}$. %Basic tower $\Bc_j$ corresponds to edge $j$. 
According to Condition \ref{con:wac:1} in Definition \ref{def:wac:cic}, and Lemma \ref{lem:decoding:Bi}, we have
\begin{align}
R_{i(1)}+R_{i(m+1)}=g(i(1),i(m+1)).  \label{eq:wac:cic:ends}
\end{align}

For the basic tower $\Bc_j$, $j\in [m]$, by Lemma \ref{lem:basic:tower} we have
\begin{align}
1\ge g(i(j),i(j+1))+\sum_{\ell \in [h_j]}R_{k_{\ell}(j)}.   \label{eq:wac:cic:eachj}
\end{align}

Summing up \eqref{eq:wac:cic:eachj} for all $j\in [m]$, we obtain
\begin{align}
m&\ge \sum_{j\in [m]}g(i(j),i(j+1))+\sum_{j\in [m]}\sum_{\ell \in [h_j]}R_{k_{\ell}(j)}    \nonumber   \\
  &=\sum_{j\in [m]}g(i(j),i(j+1))+|\Kc|R_{\rm sym}. \label{eq:wac:cic:vertical}
\end{align}

We can further bound $\sum_{j\in [m]}g(i(j),i(j+1))$ as 
\begin{align}
\sum_{j\in [m]}g(i(j),i(j+1))    
&\ge \sum_{j\in [2:m]}R_{i(j)}+g(i(1),i(m+1))   \nonumber    \\
&=|\Ic|R_{\rm sym}.    \label{eq:wac:cic:horizontal}
\end{align}
where the inequality is due to Lemma \ref{lem:horizontal}, and the equality is due to \eqref{eq:wac:cic:ends}. 

Combining \eqref{eq:wac:cic:vertical} and \eqref{eq:wac:cic:horizontal} yields $$R_{\rm sym}\le \frac{m}{|\Ic|+|\Kc|}=\frac{m}{1+m+\sum_{j\in [m]}h_j},$$ which completes the proof. 
\end{IEEEproof}

%%%%%%%%%%%%%%%%%%%%%%%%%%%%%%%%%%%%%%%%%%%%
%%%%%%%%%%%%%%%%%%%%%%%%%%%%%%%%%%%%%%%%%%%%
%%%%%%%%%%%%%%%%%%%%%%%%%%%%%%%%%%%%%%%%%%%%

%%%%%%%%%%%%%%%%%%%%%%%%%%%%%%%%%%%%%%%%%%%%
%%%%%%%%%%%%%%%%%%%%%%%%%%%%%%%%%%%%%%%%%%%%
%%%%%%%%%%%%%%%%%%%%%%%%%%%%%%%%%%%%%%%%%%%%
%THEOREM 2

%\iffalse

%\newpage

%\twocolumn
%\vspace{2mm}
\section{Proof of Theorem \ref{thm:gwac:cic}} \label{sec:modular:gwac:cic}

We first present the following lemma for the crossing tower.

\begin{lemma}   \label{lem:crossing:tower}
For the CIC problem $(i|A_i)$, $i\in [n]$ with the crossing tower $\Xcal_j$ of central edge $j$, 
we have
\begin{align}
&\sum_{j'\in G_j}\sum_{\ell\in [h_{j'}]}R_{k_{\ell}(j')}+\sum_{j' \in [s_{h_j,j}+1:t_{h_j,j}-1]}R_{i(j')}    \nonumber   \\
&\quad +g(i(s_{h_j,j}),i(t_{h_j,j})) \le |G_j|.   \label{eq:lem:crossing:tower}
\end{align}
\end{lemma}

\begin{IEEEproof}
%Consider any crossing tower $\Xcal_j$, $j\in M$. 
According to Condition \ref{con:x:tower:1} in Definiton \ref{def:x:tower}, for any $j'\in G_j\setminus \{j\}$, the message group $\{ i(j'),i(j'+1),K_{[h_{j'}]}(j') \}$ forms a basic tower $\Bc_{j'}$. 
Hence by Lemma \ref{lem:basic:tower} we have
\begin{align}
&\sum_{j'\in G_j\setminus \{j\}}\big( \sum_{\ell \in [h_{j'}]}R_{k_{\ell}(j')}+g(i(j'),i(j'+1)) \big)    \nonumber   \\
&\le \sum_{j'\in G_j\setminus \{j\}} 1=|G_j|-1.    \label{eq:lem:crossing:tower:j'}
\end{align}

For the core of $\Xcal_j$ constituted by the message group $\{ i(j),i(j+1),K_{[h_j]}(j) \}$, consider any $\ell \in [2:h_j]$, we write %using the submodularity and monotonicity of $g(S)$, we write
\begin{align}
&g(K_{[\ell-2]}(j),i(s_{\ell-1,j}),i(t_{\ell-1,j}))+R_{k_{\ell-1}(j)}    \nonumber   \\
&=g(K_{[\ell-1]}(j),i(s_{\ell-1,j}),i(t_{\ell-1,j}))     \nonumber    \\
&\le g(K_{[\ell-1]}(j),i(s_{\ell-1,j}),i(t_{\ell-1,j}), i(s_{\ell,j}),i(t_{\ell,j}))     \nonumber\\
&\le g(K_{[\ell-1]}(j),i(s_{\ell,j}),i(t_{\ell,j}))-g(i(s_{\ell,j}),i(t_{\ell,j}))         \nonumber \\   
&\quad +g(i(s_{\ell,j}),i(t_{\ell,j}),i(s_{\ell-1,j}),i(t_{\ell-1,j})),           \label{eq:lem:crossing:tower:s0}
\end{align}
where the equality follows from Lemma \ref{lem:decoding:Bi} and Conditions \ref{con:x:tower:2} and \ref{con:x:tower:31} in Definition \ref{def:x:tower}, the first inequality follows from the monotonicity of $g(S)$, and the second inequality follows from the submodularity of $g(S)$. Also by Lemma \ref{lem:decoding:Bi} and Conditions \ref{con:x:tower:2} and \ref{con:x:tower:31} in Definition \ref{def:x:tower}, as well as that $g(K_{[h_j]}(j),i(s_{h_j,j}),i(t_{h_j,j})) \le 1$, we have
\begin{align}
R_{k_{h_j}(j)}+g(K_{[h_j-1]}(j),i(s_{h_j,j}),i(t_{h_j,j}))\le 1.  \label{eq:lem:crossing:k:top}
\end{align}
Summing up \eqref{eq:lem:crossing:tower:s0} for all $\ell \in [2:h_j]$ and adding \eqref{eq:lem:crossing:k:top}, and then eliminating redundant terms on the LHS and RHS of the inequality gives
\begin{align}
&\sum_{\ell \in [h_j]}g( i(s_{\ell,j}),i(t_{\ell,j}) )+\sum_{\ell \in [h_j]}R_{k_{\ell}(j)}     \nonumber    \\
&\le 1+\sum_{\ell \in [2:h_j]}g(i(s_{\ell,j}),i(t_{\ell,j}),i(s_{\ell-1,j}),i(t_{\ell-1,j})).       \label{eq:lem:crossing:tower:s1}
\end{align}

Again consider any $\ell \in [2:h_j]$. By Lemma \ref{lem:horizontal}, we have
\begin{align}
&\sum_{j'\in [s_{\ell,j}:s_{\ell-1,j}-1]}g( i(j'),i(j'+1))       \nonumber   \\
&\ge \sum_{j' \in [s_{\ell,j}+1:s_{\ell-1,j}-1]}R_{i(j')}+g(i(s_{\ell,j}),i(s_{\ell-1,j})),     \nonumber  %\label{eq:lem:crossing:tower:s2:left}
\end{align}
and
\begin{align}
&\sum_{j'\in [t_{\ell-1,j}:t_{\ell,j}-1]}g( i(j'),i(j'+1) )       \nonumber     \\
&\ge \sum_{j' \in [t_{\ell-1,j}+1:t_{\ell,j}-1]}R_{i(j')}+g(i(t_{\ell-1,j}),i(t_{\ell,j})).         \nonumber   %\label{eq:lem:crossing:tower:s2:right}
\end{align}
Also, we have
\begin{align*}
&g(i(s_{\ell,j}),i(s_{\ell-1,j}))+g( i(s_{\ell-1,j}),i(t_{\ell-1,j}) )      \\
&\quad +g(i(t_{\ell-1,j}),i(t_{\ell,j}))    \\
&\ge R_{i(s_{\ell-1,j})}+R_{i(t_{\ell-1,j})}    \\
&\quad +g( i(s_{\ell,j}),i(s_{\ell-1,j}),i(t_{\ell-1,j}),i(t_{\ell,j}) ), 
\end{align*}
due to the submodularity of $g(S)$ as well as Lemma \ref{lem:decoding:Bi}. 
Combinning the above three inequalities, we have
\begin{align}
&\sum_{j'\in [s_{\ell,j}:s_{\ell-1,j}-1]}g( i(j'),i(j'+1))+g( i(s_{\ell-1,j}),i(t_{\ell-1,j}) )    \nonumber    \\
&\quad +\sum_{j'\in [t_{\ell-1,j}:t_{\ell,j}-1]}g( i(j'),i(j'+1) )   \nonumber     \\
&\geq  \sum_{j' \in [s_{\ell,j}+1:s_{\ell-1,j}]}R_{i(j')}+\sum_{j' \in [t_{\ell-1,j}:t_{\ell,j}-1]}R_{i(j')}    \nonumber    \\
&\quad +g( i(s_{\ell,j}),i(s_{\ell-1,j}),i(t_{\ell-1,j}),i(t_{\ell,j}) ).  \label{eq:lem:crossing:tower:s2}
\end{align}

Condition \ref{con:x:tower:32} in Definition \ref{def:x:tower} states that for any $\ell \in [2:h_j]$, $s_{\ell,j}\le s_{\ell-1,j}\le s_{1,j}=j$, and $j+1=t_{1,j}=t_{\ell-1,j}\le t_{\ell,j}$. 
Hence, adding up \eqref{eq:lem:crossing:tower:s2} for all $\ell \in [2:h_j]$ and rearranging yields 
\iffalse
indicates that for any $\ell_1 \neq \ell_2 \in [2:h_j]$, 
sets $ [s_{\ell_1,j}:s_{\ell_1-1,j}-1] $, $[s_{\ell_2,j}:s_{\ell_2-1,j}-1]$, $[t_{\ell_1-1,j}:t_{\ell_1,j}-1]$, and $[t_{\ell_2-1,j}:t_{\ell_2,j}-1]$ are mutually disjoint. 
So do sets $ [s_{\ell_1,j}+1:s_{\ell_1-1,j}] $, $[s_{\ell_2,j}+1:s_{\ell_2-1,j}]$, $[t_{\ell_1-1,j}:t_{\ell_1,j}-1]$, and $[t_{\ell_2-1,j}:t_{\ell_2,j}-1]$. 
Also, recall that $s_{j,1}=j$, and $t_{j,1}=j+1$. 
Therefore, summing up \eqref{eq:lem:crossing:tower:s2} for all $\ell \in [2:h_j]$ and then multiplying by $-1$ on both sides gives 
\fi
\begin{align}
&\sum_{j' \in [s_{h_j,j}+1:t_{h_j,j}-1]}R_{i(j')}-\sum_{j'\in G_j\setminus \{j\}}g(i(j'),i(j'+1))    \nonumber    \\
&\quad -\sum_{\ell \in [2:h_j]}g(i(s_{\ell-1,j}),i(t_{\ell-1,j}))   \nonumber   \\
&\le -\sum_{\ell \in [2:h_j]}g( i(s_{\ell,j}),i(s_{\ell-1,j}),i(t_{\ell-1,j}),i(t_{\ell,j}) ).   \label{eq:lem:crossing:tower:s4}
\end{align}

Finally, summing up \eqref{eq:lem:crossing:tower:j'}, \eqref{eq:lem:crossing:tower:s1}, and \eqref{eq:lem:crossing:tower:s4} and simplifying the result yields \eqref{eq:lem:crossing:tower}, which completes the proof.
\end{IEEEproof}

We prove Theorem \ref{thm:gwac:cic} in the following with the help of Lemmas \ref{lem:decoding:Bi}, \ref{lem:basic:tower}, and \ref{lem:crossing:tower}. 

\begin{IEEEproof}
Consider the disjoint weighted alignment chain $i(1)\xleftrightarrow{\Ic,\Kcal } i(m+1)$ for a CIC problem. % for the CIC problem under consideration, 
By Lemma \ref{lem:decoding:Bi} and Condition \ref{con:gwac:1} in Definition \ref{def:gwac:cic}, we have
\begin{align}
R_{i(1)}+R_{i(m+1)}=g( i(1),i(m+1) ).           \label{eq:gwac:cic:modular:ends}
\end{align}

Recall that any disjoint weighted alignment chain can be seen as a concatenation of the crossing towers $\Xcal_j$, $j\in M=\{ j\in [m]:|G_j|\ge 2 \}$ and the basic towers $\Bc_{j'}$, $j'\in M'=[m]\setminus (\bigcup_{j\in M}G_{j})$. Hence, by Lemmas  \ref{lem:basic:tower} and \ref{lem:crossing:tower}, we have
\begin{align}
&\sum_{j' \in M'}\big( \sum_{\ell \in [h_{j'}]}R_{k_{\ell}(j')}+g( i(j'),i(j'+1) ) \big)     \nonumber    \\
&\quad +\sum_{j\in M}\big( \sum_{j'\in G_j}\sum_{\ell\in [h_{j'}]}R_{k_{\ell}(j')}+\sum_{j' \in [s_{h_j,j}+1:t_{h_j,j}-1]}R_{i(j')}   \nonumber   \\
&\quad +g(i(s_{h_j,j}),i(t_{h_j,j})) \big)\le \sum_{j'\in M'}1+\sum_{j\in M}|G_j|.   \label{eq:gwac:cic:modular:final}
\end{align}

We further bound the LHS and the RHS of \eqref{eq:gwac:cic:modular:final} in the following. 

For the RHS of \eqref{eq:gwac:cic:modular:final}, we have
\begin{align}
RHS&=\sum_{j\in M}|G_j|+|[m]\setminus (\bigcup_{j\in M}G_j)|=m.    \label{eq:gwac:cic:modular:final:RHS}
\end{align}

Similarly, for the LHS of \eqref{eq:gwac:cic:modular:final}, we have
\begin{align}
&LHS
%=\sum_{j' \in M'}\sum_{\ell \in [h_{j'}]}R_{k_{\ell}(j')}+\sum_{j\in M}\sum_{j'\in G_j}\sum_{\ell\in [h_{j'}]}R_{k_{\ell}(j')}  \nonumber   \\
%&\quad +\sum_{j\in M}\sum_{j' \in [s_{h_j,j}+1:t_{h_j,j}-1]}R_{i(j')}               \nonumber   \\
%&\quad +\sum_{j\in M}g( i(s_{h_j,j}),i(t_{h_j,j}) )+\sum_{j' \in M'}g( i(j'),i(j'+1) )       \nonumber     \\
= \sum_{j\in [m]}\sum_{\ell\in [h_j]}R_{k_{\ell}(j)}+\sum_{j\in M}\sum_{j' \in [s_{h_j,j}+1:t_{h_j,j}-1]}R_{i(j')}               \nonumber   \\
&+\sum_{j\in M}g( i(s_{h_j,j}),i(t_{h_j,j}) )+\sum_{j' \in M'}g( i(j'),i(j'+1) ).   \label{eq:gwac:cic:modular:final:LHS:mid:0}
\end{align}

As any basic tower $\Bc_{j'}$, $j'\in M'$ is a special crossing tower with $s_{h_{j'},j'}=j'$, $t_{h_{j'},j'}=j'+1$, we have
\begin{align}
&\sum_{j\in M}g( i(s_{h_j,j}),i(t_{h_j,j}) )+\sum_{j' \in M'}g( i(j'),i(j'+1) )    \nonumber   \\
&=\sum_{j\in M\cup M'}g( i(s_{h_j,j}),i(t_{h_j,j}) ).      \label{eq:gwac:cic:modular:final:LHS:mid:1}
\end{align}
Note that the set $M\cup M'$ denotes the collection of central edges of the crossing towers $\Xcal_j$, $j\in M$ and the basic towers $\Bc_{j'}$, $j'\in M'$.  
As these towers are concatenated to form the chain, if we order the elements of the set $M\cup M'$ as $j_1<j_2<\cdots<j_{|M\cup M'|}$, then we have
\begin{align}
&s_{h_{j_1},j_1}=1,         \label{eq:gwac:cic:order:left:end}   \\
&t_{h_{j_{|M\cup M'|}},j_{|M\cup M'|}}=m+1,             \label{eq:gwac:cic:order:right:end}    \\
&s_{h_{j_{p}},j_{p}}=t_{h_{j_{p-1}},j_{p-1}},  \qquad \quad \forall p \in [2:|M\cup M'|].   \label{eq:gwac:cic:order:dula}
\end{align}
For any $p \in [2:|M\cup M'|]$, by the submodularity and the monotonicity of $g(S)$, as well as Lemma \ref{lem:decoding:Bi}, we have
\begin{align}
&g(i(s_{h_{j_1},j_1}),i(s_{h_{j_{p}},j_{p}}))+g(i(s_{h_{j_{p}},j_{p}}), i(t_{h_{j_{p}},j_{p}}))            \nonumber    \\
&\ge R_{i(s_{h_{j_{p}},j_{p}})}+g(i(s_{h_{j_1},j_1}),i(t_{h_{j_{p}},j_{p}})).     \nonumber   
\end{align}
Given the fact that $j_p$, $p\in [|M\cup M'|]$ is a reindexing of $j$, $j\in M\cup M'$, as well as \eqref{eq:gwac:cic:order:left:end}-\eqref{eq:gwac:cic:order:dula}, summing up the above inequality for all $p \in [2:|M\cup M'|]$ and simplifying yields
\begin{align}
&\sum_{j\in M\cup M'}g( i(s_{h_j,j}),i(t_{h_j,j}) )    \nonumber    \\
&\ge \sum_{p \in [2:|M\cup M'|]}R_{i(s_{h_{j_{p}},j_{p}})}+g(i(s_{h_{j_1},j_1}),i(t_{h_{j_{|M\cup M'|}},j_{|M\cup M'|}}))    \nonumber  \\
&=\sum_{j' \in (\bigcup_{j\in M\cup M'}\{ s_{h_j,j},t_{h_j,j} \})\setminus \{ 1,m+1 \}}R_{i(j')}+g(i(1),i(m+1))    \nonumber   \\
&=\sum_{j' \in \bigcup_{j\in M\cup M'}\{ s_{h_j,j},t_{h_j,j} \}}R_{i(j')}, \label{eq:gwac:cic:modular:final:LHS:mid:2} 
\end{align}
where the last equality is due to \eqref{eq:gwac:cic:modular:ends}. 

Combinning \eqref{eq:gwac:cic:modular:final:LHS:mid:0}, \eqref{eq:gwac:cic:modular:final:LHS:mid:1} and \eqref{eq:gwac:cic:modular:final:LHS:mid:2}, we can bound the LHS of \eqref{eq:gwac:cic:modular:final} as
\begin{align}
LHS&\ge \sum_{j\in [m]}\sum_{\ell\in [h_j]}R_{k_{\ell}(j)}+\sum_{j\in M}\sum_{j' \in [s_{h_j,j}+1:t_{h_j,j}-1]}R_{i(j')}    \nonumber   \\
&\quad +\sum_{j' \in \bigcup_{j\in M\cup M'}\{ s_{h_j,j},t_{h_j,j} \}}R_{i(j')}     \nonumber    \\
&=\sum_{j\in [m]}\sum_{\ell\in [h_j]}R_{k_{\ell}(j)}+\sum_{j\in [m+1]}R_{i(j)}   \nonumber  \\
&=(|\Kcal|+|\Ic|)R_{\rm sym}.    \label{eq:gwac:cic:modular:final:LHS}
\end{align}

Given \eqref{eq:gwac:cic:modular:final}, \eqref{eq:gwac:cic:modular:final:RHS}, and \eqref{eq:gwac:cic:modular:final:LHS}, we can conclude that
\begin{align}
R_{\rm sym}\le \frac{m}{|\Ic|+|\Kcal|}=\frac{m}{1+m+\sum_{j\in [m]}h_j},
\end{align}
which completes the proof. 
\end{IEEEproof}

%\fi

%%%%%%%%%%%%%%%%%%%%%%%%%%%%%%%%%%%%%%%%%%%%
%%%%%%%%%%%%%%%%%%%%%%%%%%%%%%%%%%%%%%%%%%%%
%%%%%%%%%%%%%%%%%%%%%%%%%%%%%%%%%%%%%%%%%%%%

%%%%%%%%%%%%%%%%%%%%%%%%%%%%%%%%%%%%%%%%%%%%
%%%%%%%%%%%%%%%%%%%%%%%%%%%%%%%%%%%%%%%%%%%%
%%%%%%%%%%%%%%%%%%%%%%%%%%%%%%%%%%%%%%%%%%%%
%THEOREM 3

%\newpage

%\vspace{2mm}
\section{Proof of Theorem \ref{thm:wac:dic}} \label{sec:proof:wac:dic}

\begin{lemma}   \label{lem:basic:tower:dic}
For the DIC problem $(i|A_i)$, $i\in [n]$ with the basic tower $\Bc_j$ constituted by the message group $\{ i(j),i(j+1),K_{[h_j]}(j) \}$, 
we have
\begin{align}
&f([n]; i(j),i(j+1) )+\sum_{\ell \in [h_j]}R_{k_{\ell}(j)}    \nonumber   \\
&\le f([n]; i(j),i(j+1),K_{[h_j]}(j) ).   \label{eq:lem:basic:tower:dic}
\end{align}
\end{lemma}
Similar to Lemma \ref{lem:basic:tower} which is shown via repeated application of Lemma \ref{lem:decoding:Bi}, 
Lemma \ref{lem:basic:tower:dic} above can be shown via repeated application of Lemma \ref{lem:decoding:Bi:dic}. 

\begin{lemma}   \label{lem:wac:linking}
For the DIC problem $(i|A_i)$, $i\in [n]$ with the weighted alignment chain $i(1)\xleftrightarrow{\Ic,\Kcal } i(m+1)$, for any $a,b,c,d\in [m]$, we have
\begin{align}
&f(i(a),i(b),K_{[h_d]}(d) ; i(a),i(c),K_{[h_d]}(d) )       \nonumber     \\
&\quad +f([n]; i(a),i(b),i(c) )   \nonumber    \\
&\ge f([n]; i(a),i(b),K_{[h_d]}(d) )+f([n]; i(a),i(c) ).  \label{eq:wac:linking}
\end{align}
\end{lemma}
%The above lemma can be shown using the submodularity and the monotonicity and property \eqref{eq:f:linking} of $f(L,S)$. 
\begin{IEEEproof}
We have
\begin{align*}
&f (i(a),i(b),K_{[h_d]}(d) ; i(a),i(c),K_{[h_d]}(d) )       \\
&\quad +f([n]; i(a),i(b),i(c) )     \\
&\ge f (i(a),i(b),K_{[h_d]}(d) ; i(a),i(b),i(c),K_{[h_d]}(d))   \\
&\quad +f([n]; i(a),i(c) )   \\
&\ge f (i(a),i(b),K_{[h_d]}(d) ; i(a),i(b),K_{[h_d]}(d))   \\
&\quad +f([n]; i(a),i(c) )   \\
&=f([n]; i(a),i(b),K_{[h_d]}(d) )+f([n]; i(a),i(c) ),
\end{align*}
where the first and second inequalities are due to the submodularity and monotonicity of $f(L;S)$, respectively, and the equality is due to property \eqref{eq:f:linking} of $f(L;S)$. 
\end{IEEEproof}

%Using Lemmas \ref{lem:decoding:Bi:dic} and \ref{lem:wac:linking}, we can show the following lemma,
\begin{lemma}     \label{lem:horizontal:dic}
For the DIC problem $(i|A_i)$, $i\in [n]$ with the weighted alignment chain $i(1)\xleftrightarrow{\Ic,\Kcal } i(m+1)$, for any two integers $a<b\in [m]$, we have
\begin{align}
&\sum_{j\in [a:b]}f([n]; i(j),i(j+1) )          \nonumber    \\
&\ge \sum_{j\in [a+1:b]}R_{i(j)}+f([n]; i(a),i(b+1) )     \nonumber     \\
&\quad +\sum_{j\in [a:b]}f([n]; i(j),i(j+1),i(b+1) )     \nonumber   \\
&\quad -\sum_{j\in [a:b]}f([n]; i(j),i(b+1) ).    \label{eq:lem:horizontal:dic}
\end{align}
\end{lemma}

\begin{IEEEproof}
Due to the fact that when $j=b$, $f([n]; i(j),i(b+1) )=f([n]; i(j),i(j+1),i(b+1) )$, it suffices to show that
\begin{align}
&\sum_{j\in [a:b]}f([n]; i(j),i(j+1) )          \nonumber    \\
&\ge \sum_{j\in [a+1:b]}R_{i(j)}+f([n]; i(a),i(b+1) )     \nonumber     \\
&\quad +\sum_{j\in [a:b-1]}f([n]; i(j),i(j+1),i(b+1) )     \nonumber   \\
&\quad -\sum_{j\in [a:b-1]}f([n]; i(j),i(b+1) ).    \label{eq:lem:horizontal:dic:1}
\end{align}

For any $j\in [a:b-1]$, by the submodularity of $f(L;S)$ and Lemma \ref{lem:decoding:Bi:dic}, we have
\begin{align}
&f([n]; i(b+1),i(j+1) )+f([n]; i(j+1),i(j) )      \nonumber   \\
&\ge R_{i(j+1)}+f([n]; i(b+1),i(j) )       \nonumber    \\
&\quad +f([n]; i(b+1),i(j+1),i(j) )       \nonumber    \\
&\quad -f([n]; i(b+1),i(j) ).       \label{eq:lem:horizontal:dic:j}
\end{align}

Summing up \eqref{eq:lem:horizontal:dic:j} for all $j\in [a:b-1]$ and simplifying the result yields \eqref{eq:lem:horizontal:dic:1}, and thus completes the proof. 
\end{IEEEproof}

%\iffalse
%\vspace{10mm}

%\textcolor{red}{tbc}

We show Theorem \ref{thm:wac:dic} as follows.

\iffalse
\begin{theorem}
Given any DIC problem $(i|A_i)$, $i\in [n]$, if $R_{\rm sym}$ is achievable, then for any singleton weighted alignment chain $i(1)\xleftrightarrow{\Ic,\Kcal }_{\rm s} i(m+1)$ we have
\begin{align*}
R_{\rm sym}\leq &\frac{1}{|\Ic|+|\Kcal|}\sum_{j\in [m]}\sum_{\substack{J\cap \{ i(j),i(j+1),K_{[h_{j+1}]}(j+1)\} \neq \emptyset, \\ J\cap \{ i(j),i(m+1),K_{[h_{j+1}]}(j+1)\} \neq \emptyset}}C_J. 
\end{align*}
Goal:
\begin{align*}
&(|\Ic|+|\Kcal|)R_{\rm sym}   \nonumber   \\
&\le \sum_{j\in [m]}f(\{ i(j),i(m+1),K_{[h_j]}(j) \},\{ i(j),i(j+1),K_{[h_j]}(j) \})
\end{align*}
\end{theorem}
\fi

\begin{IEEEproof}
Consider the singleton weighted alignment chain $i(1)\xleftrightarrow{\Ic,\Kc}_{\rm s} i(m+1)$ for a DIC problem. According to Condition \ref{con:wac:1} in Definition \ref{def:wac:cic} and Lemma \ref{lem:decoding:Bi:dic}, we have
\begin{align}
f([n]; i(m+1),i(1) )=R_{i(m+1)}+R_{i(1)}.   \label{eq:wac:dic:terminals}
\end{align}

As the singleton weighted alignment chain can be seen as a concatenation of basic towers $\Bc_j$, $j\in [m]$, by Lemma \ref{lem:basic:tower:dic} we have
\begin{align}
&\sum_{j\in [m]}f([n]; i(j),i(j+1) )+\sum_{j\in [m]}\sum_{\ell \in [h_j]}R_{k_{\ell}(j)}   \nonumber   \\
&\le \sum_{j\in [m]}f([n]; i(j),i(j+1),K_{[h_j]}(j) ).    \label{eq:wac:dic:all:basic:tower}
\end{align}

By Lemma \ref{lem:horizontal:dic}, we can bound the LHS of \eqref{eq:wac:dic:all:basic:tower} as 
\begin{align}
LHS&=\sum_{j\in [m]}f([n]; i(j),i(j+1) )+|\Kcal|R_{\rm sym}   \nonumber   \\
&\ge \sum_{j\in [2:m]}R_{i(j)}+f([n]; i(1),i(m+1) )     \nonumber     \\
&\quad +\sum_{j\in [m]}f([n]; i(j),i(j+1),i(m+1) )     \nonumber   \\
&\quad -\sum_{j\in [m]}f([n]; i(j),i(m+1) )+|\Kcal|R_{\rm sym}         \nonumber     \\
&=(|\Ic|+|\Kcal|)R_{\rm sym}    \nonumber    \\
&\quad +\sum_{j\in [m]}f([n]; i(m+1),i(j+1),i(j) )    \nonumber   \\
&\quad -\sum_{j\in [m]}f([n]; i(m+1),i(j) ),            \label{eq:wac:dic:LHS}
\end{align}
where the last equality follows from \eqref{eq:wac:dic:terminals}.

Combinning \eqref{eq:wac:dic:all:basic:tower} and \eqref{eq:wac:dic:LHS}, we have
\begin{align}
&(|\Ic|+|\Kcal|)R_{\rm sym}   \nonumber   \\
&\le  \sum_{j\in [m]}f([n]; i(j),i(j+1),K_{[h_j]}(j) )       \nonumber   \\
&\quad -\sum_{j\in [m]}f([n]; i(m+1),i(j+1),i(j) )    \nonumber   \\
&\quad +\sum_{j\in [m]}f([n]; i(m+1),i(j) )           \nonumber     \\
&\le \sum_{j\in [m]}f( i(j),i(j+1),K_{[h_j]}(j) ; i(j),i(m+1),K_{[h_j]}(j) )      \nonumber   \\
&\le \sum_{j\in [m]}\sum_{\substack{J\in N: J\cap \{ i(j),i(j+1),K_{[h_j]}(j)\} \neq \emptyset, \\ J\cap \{ i(j),i(m+1),K_{[h_j]}(j)\} \neq \emptyset}}C_J. 
\end{align}
where %the first inequality follows from \eqref{eq:wac:dic:all:basic:tower} and \eqref{eq:wac:dic:LHS:mid}, and 
the second inequality follows from Lemma \ref{lem:wac:linking} with $a=d=j$, $b=j+1$, and $c=m+1$ for any $j\in [m]$, and the last inequality follows from property \eqref{eq:f:capacity:constraint} of $f(L;S)$. 
\end{IEEEproof}

%\fi

%%%%%%%%%%%%%%%%%%%%%%%%%%%%%%%%%%%%%%%%%%%%
%%%%%%%%%%%%%%%%%%%%%%%%%%%%%%%%%%%%%%%%%%%%
%%%%%%%%%%%%%%%%%%%%%%%%%%%%%%%%%%%%%%%%%%%%
%THEOREM 4

%\twocolumn
%\newpage
%\vspace{2mm}
\section{Proof of Theorem 4} \label{sec:proof:gwac:dic}

For easier reference, we repeat \eqref{eq:gwac:dic} in Theorem \ref{thm:gwac:dic} below,
\iffalse 
\begin{align}
&R_{\rm sym}\leq 
\frac{1}{|\Ic|+|\Kcal|}\big(
\sum_{j\in M}\sum_{\ell \in [2:h_j]}( \sum_{j' \in [s_{\ell,j}:s_{\ell-1,j}-1]}\sum_{\substack{J\cap T_3\neq \emptyset,\\ J\cap T_4\neq \emptyset}}C_J       \nonumber  \\ 
&+\sum_{j' \in [t_{\ell-1,j}:t_{\ell,j}-1]}\sum_{\substack{J\cap T_3\neq \emptyset,\\ J\cap T_5\neq \emptyset}}C_J )+\sum_{\substack{j\in \\M\cup M'}}\sum_{\substack{J\cap T_1\neq \emptyset,\\ J\cap T_2\neq \emptyset}}C_J  \big),  \label{eq:gwac:dic:repeat}
\end{align}
\fi
\begin{align}
R_{\rm sym}   
\le &\frac{1}{|\Ic|+|\Kcal|}\big(
\sum_{\substack{j\in M\cup M'}}\sum_{\substack{J\in N: J\cap T_1(j)\neq \emptyset, J\cap T_2(j)\neq \emptyset}}C_J   \nonumber   \\
&+\sum_{j\in M}\sum_{\ell \in [2:h_j]}( \sum_{j' \in [s_{\ell,j}:s_{\ell-1,j}-1]}\sum_{\substack{J\in N: \\ J\cap T_3(j')\neq \emptyset,\\ J\cap T_4(j,\ell,j')\neq \emptyset}}C_J       \nonumber  \\ 
&+\sum_{j' \in [t_{\ell-1,j}:t_{\ell,j}-1]}\sum_{\substack{J\in N: \\ J\cap T_3(j')\neq \emptyset,\\ J\cap T_5(j,\ell,j')\neq \emptyset}}C_J ) \big),  \label{eq:gwac:dic:repeat}
\end{align}
where 
\begin{align*}
&T_1(j)=\{ i(s_{h_j,j}),i(t_{h_j,j}),K_{[h_j]}(j) \},   \nonumber    \\
&T_2(j)=\{ i(s_{h_j,j}),i(m+1),K_{[h_j]}(j) \},      \nonumber     \\
&T_3(j')=\{ i(j'),i(j'+1),K_{[h_{j'}]}(j') \},   \nonumber   \\
&T_4(j,\ell,j')=\{ i(j'),i(s_{\ell-1,j}),K_{[h_{j'}]}(j') \},     \nonumber    \\
&T_5(j,\ell,j')=\{ i(j'),i(t_{\ell,j}),K_{[h_{j'}]}(j') \}.
\end{align*}

We first show the following lemma for the crossing tower. % in a DIC problem.

\begin{lemma}    \label{lem:x:tower:dic}
For the DIC problem $(i|A_i)$, $i\in [n]$ with link capacity tuple $\Cv$ and the crossing tower $\Xcal_j$ of central edge $j$, we have
\begin{align}
&\sum_{j'\in G_j}\sum_{\ell\in [h_{j'}]}R_{k_{\ell}(j')}     \nonumber  \\
&\quad +f([n]; i(s_{h_j,j}),i(t_{h_j,j}) )+\sum_{\ell \in [s_{h_j,j}+1:t_{h_j,j}-1]}R_{i(\ell)}    \nonumber   \\
%%%
&\le f([n]; K_{[h_j]}(j),i(s_{h_j,j}),i(t_{h_j,j}) )   \nonumber  \\
&\quad +\sum_{\ell \in [2:h_j]}\sum_{j' \in [s_{\ell,j}:s_{\ell-1,j}-1]}\sum_{\substack{J\in N: J\cap T_3(j')\neq \emptyset, \\ J\cap T_4(j,\ell,j')\neq \emptyset}}C_J  \nonumber  \\
&\quad +\sum_{\ell \in [2:h_j]}\sum_{j' \in [t_{\ell-1,j}:t_{\ell,j}-1]}\sum_{\substack{J\in N: J\cap T_3(j')\neq \emptyset, \\ J\cap T_5(j,\ell,j')\neq \emptyset}}C_J.  \label{eq:lem:crossing:tower:dic:goal}
\end{align}
\end{lemma}

\begin{IEEEproof}
Within the crossing tower $\Xcal_j$, any edge $j'\in G_{j}\setminus \{j\}$ corresponds to a basic tower $\Bc_{j'}$. Thus by Lemma \ref{lem:basic:tower:dic} we have
\begin{align}
&\sum_{j'\in G_j\setminus \{j\}}\big( \sum_{\ell \in [h_{j'}]}R_{k_{\ell}(j')}+f([n]; i(j'),i(j'+1) ) \big)    \nonumber   \\
&\le \sum_{j'\in G_j\setminus \{j\}} f([n]; i(j'),i(j'+1),K_{[h_{j'}]}(j') ).    \label{eq:lem:crossing:tower:j':dic}
\end{align}

%%%^^^^^^^^^^^^^^^^^^^^
Consider any $\ell\in [2:h_j]$. We have
\begin{align}
&f([n]; K_{[\ell-1]}(j),i(s_{\ell,j}),i(t_{\ell,j}) )-f([n]; i(s_{\ell,j}),i(t_{\ell,j}) )     \nonumber    \\
&\quad +f([n]; i(s_{\ell,j}),i(t_{\ell,j}),i(s_{\ell-1,j}),i(t_{\ell-1,j}) )     \nonumber \\
&\ge f([n]; K_{[\ell-1]}(j),i(s_{\ell-1,j}),i(t_{\ell-1,j}) )   \nonumber    \\
&= f([n]; K_{[\ell-2]}(j),i(s_{\ell-1,j}),i(t_{\ell-1,j}) )+R_{k_{\ell-1}(j)},       \label{eq:lem:crossing:tower:core1:dic}
\end{align}
where the inequality follows from the submodularity and the monotonicity of $f(L;S)$, and the equality follows from Lemma \ref{lem:decoding:Bi:dic} with $B_{k_{\ell-1}(j)}\supseteq \{ K_{[\ell-2]}(j),i(s_{\ell-1,j}),i(t_{\ell-1,j}) \}$ by Definition \ref{def:gwac:cic}. Summing up \eqref{eq:lem:crossing:tower:core1:dic} for all $\ell \in [2:h_j]$ and removing redundant terms yields
\begin{align}
&f([n]; K_{[h_j-1]}(j),i(s_{h_j,j}),i(t_{h_j,j}) )   \nonumber   \\
&\quad +\sum_{\ell \in [2:h_j]}f([n]; i(s_{\ell,j}),i(t_{\ell,j}),i(s_{\ell-1,j}),i(t_{\ell-1,j}) )    \nonumber   \\
&\ge \sum_{\ell \in [h_j]}f([n]; i(s_{\ell,j}),i(t_{\ell,j}) )+\sum_{\ell \in [h_j-1]}R_{k_{\ell}(j)}.        \label{eq:lem:crossing:tower:core2:dic}
\end{align}
Thus, we have 
\begin{align}
&  \sum_{\ell \in [h_j]}f([n]; i(s_{\ell,j}),i(t_{\ell,j}) )+\sum_{\ell \in [h_j]}R_{k_{\ell}(j)}   \nonumber   \\
&\le \sum_{\ell \in [2:h_j]}f([n]; i(s_{\ell,j}),i(t_{\ell,j}),i(s_{\ell-1,j}),i(t_{\ell-1,j}) )    \nonumber   \\
&\quad +f([n]; K_{[h_j-1]}(j),i(s_{h_j,j}),i(t_{h_j,j}) )+R_{k_{h_j}(j)}   \nonumber  \\
&=\sum_{\ell \in [2:h_j]}f([n]; i(s_{\ell,j}),i(t_{\ell,j}),i(s_{\ell-1,j}),i(t_{\ell-1,j}) )  \nonumber   \\
&\quad +f([n]; K_{[h_j]}(j),i(s_{h_j,j}),i(t_{h_j,j}) ).       \label{eq:lem:crossing:tower:core3:dic}
\end{align}
where the inequality follows from \eqref{eq:lem:crossing:tower:core2:dic} and the equality follows from Lemma \ref{lem:decoding:Bi:dic} with $B_{k_{h_j}(j)}\supseteq \{K_{[h_j-1]}(j),i(s_{h_j,j}),i(t_{h_j,j}) \}$ by Definition \ref{def:gwac:cic}.

%%%^^^^^^^^^^^^^^^^^^^^
Again consider any $\ell \in [2:h_j]$ and define shorthand notation $F_{\ell}^s$, $F_{\ell}^t$, and $F_{\ell}$ as follows, 
\begin{align*}
F_{\ell}^s &\doteq \sum_{j'=s_{\ell,j}}^{s_{\ell-1,j}-1}\sum_{\ell' \in [h_{j'}]}R_{k_{\ell'}(j')}+f([n]; i(s_{\ell,j}),i(s_{\ell-1,j}) ),        \\
F_{\ell}^t &\doteq \sum_{j'=t_{\ell-1,j}}^{t_{\ell,j}-1}\sum_{\ell' \in [h_{j'}]}R_{k_{\ell'}(j')}+f([n]; i(t_{\ell-1,j}),i(t_{\ell,j}) ),         \\
F_{\ell} &\doteq f([n]; i(s_{\ell,j}),i(s_{\ell-1,j}) )+f([n]; i(t_{\ell-1,j}),i(t_{\ell,j}) )           \\
           &\quad +f([n]; i(s_{\ell-1,j}),i(t_{\ell-1,j}) ).
\end{align*}

Note that the Condition \ref{con:x:tower:32} in Definition \ref{def:x:tower} indicates that for any $\ell_1 \neq \ell_2 \in [2:h_j]$, 
sets $ [s_{\ell_1,j}:s_{\ell_1-1,j}-1] $, $[s_{\ell_2,j}:s_{\ell_2-1,j}-1]$, $[t_{\ell_1-1,j}:t_{\ell_1,j}-1]$, and $[t_{\ell_2-1,j}:t_{\ell_2,j}-1]$ are mutually disjoint. 
Also, recall that $s_{j,1}=j$, and $t_{j,1}=j+1$. 
Hence, one can verify that 
\begin{align}
&\sum_{j'\in G_j}\sum_{\ell\in [h_{j'}]}R_{k_{\ell}(j')}+f([n]; i(s_{h_j,j}),i(t_{h_j,j}) )    \nonumber   \\
%%%
&=\sum_{\ell\in [h_j]}f([n]; i(s_{\ell,j}),i(t_{\ell,j}) )+\sum_{\ell \in [h_j]}R_{k_{\ell}(j)}    \nonumber    \\
&\quad +\sum_{\ell \in [2:h_j]}F_{\ell}^s        +\sum_{\ell \in [2:h_j]}F_{\ell}^t       -\sum_{\ell \in [2:h_j]}F_{\ell}.    \nonumber 
\end{align}
According to the above equation, to show \eqref{eq:lem:crossing:tower:dic:goal}, it suffices to show that
\begin{align}
&\sum_{\ell\in [h_j]}f([n]; i(s_{\ell,j}),i(t_{\ell,j}) )+\sum_{\ell \in [h_j]}R_{k_{\ell}(j)}    \nonumber    \\
&\quad +\sum_{\ell \in [2:h_j]}(F_{\ell}^s+F_{\ell}^t-F_{\ell})+\sum_{\ell \in [s_{h_j,j}+1:t_{h_j,j}-1]}R_{i(\ell)}         \nonumber      \\
&\le \sum_{\ell \in [2:h_j]}\sum_{j' \in [s_{\ell,j}:s_{\ell-1,j}-1]}\sum_{\substack{J\in N: J\cap T_3(j')\neq \emptyset, \\ J\cap T_4(j,\ell,j')\neq \emptyset}}C_J  \nonumber  \\
&\quad +\sum_{\ell \in [2:h_j]}\sum_{j' \in [t_{\ell-1,j}:t_{\ell,j}-1]}\sum_{\substack{J\in N: J\cap T_3(j')\neq \emptyset, \\ J\cap T_5(j,\ell,j')\neq \emptyset}}C_J    \nonumber    \\
&\quad +f([n]; K_{[h_j]}(j),i(s_{h_j,j}),i(t_{h_j,j}) ).     \label{eq:lem:crossing:tower:dic:goal:new}
%&\quad +\sum_{\ell \in [2:h_j]}\sum_{j'=t_{\ell-1,j}}^{t_{\ell,j}-1}\sum_{\substack{J\subseteq [n]: \\ J\cap T_3\neq \emptyset, \\ J\cap T_5\neq \emptyset}}C_J.
\end{align}

We bound $F_{\ell}^s$, $F_{\ell}^t$, and $F_{\ell}$ in the following. First, we have
\begin{align}
F_{\ell}^s            %\nonumber     \\
%&\sum_{j'\in [s_{\ell,j}:s_{\ell-1,j}-1]}\sum_{\ell' \in [h_{j'}]}R_{k_{\ell'}(j')}+f([n], \{ i(s_{\ell,j},i(s_{\ell-1,j})) \})               \nonumber    \\
%%
&=\sum_{j'=s_{\ell,j}}^{s_{\ell-1,j}-1}(\sum_{\ell' \in [h_{j'}]}R_{k_{\ell'}(j')}+f([n]; i(j'),i(j'+1) ))             \nonumber    \\
&\quad -\sum_{j'\in [s_{\ell,j}:s_{\ell-1,j}-1]}f([n]; i(j'),i(j'+1) )             \nonumber    \\
&\quad +f([n]; i(s_{\ell,j}),i(s_{\ell-1,j}) )                 \nonumber    \\
&\le \sum_{j'=s_{\ell,j}}^{s_{\ell-1,j}-1}f([n]; i(j'),i(j'+1),K_{[h_{j'}]}(j') )      \nonumber      \\
&\quad -\sum_{j'\in [s_{\ell,j}:s_{\ell-1,j}-1]}f([n]; i(j'),i(j'+1) )             \nonumber    \\
&\quad +f([n]; i(s_{\ell,j}),i(s_{\ell-1,j}) )                 \nonumber    \\
&\le \sum_{j'=s_{\ell,j}}^{s_{\ell-1,j}-1}f([n]; i(j'),i(j'+1),K_{[h_{j'}]}(j') )      \nonumber      \\
&\quad +\sum_{j'\in [s_{\ell,j}:s_{\ell-1,j}-1]}f([n]; i(j'),i(s_{\ell-1,j}) )     \nonumber    \\
&\quad -\sum_{j'\in [s_{\ell,j}:s_{\ell-1,j}-1]}f([n]; i(j'),i(j'+1),i(s_{\ell-1,j}) )          \nonumber    \\
&\quad -\sum_{j'\in [s_{\ell,j}+1:s_{\ell-1,j}-1]}R_{i(j')}     \nonumber    \\
%&\quad -f([n]; i(s_{\ell,j}),i(s_{\ell-1,j}) )+f([n]; i(s_{\ell,j}),i(s_{\ell-1,j}) )     \nonumber    \\
%%
&\le \sum_{j'\in [s_{\ell,j}:s_{\ell-1,j}-1]}f(T_3(j');T_4(j,\ell,j'))     \nonumber     \\
&\quad -\sum_{j'\in [s_{\ell,j}+1:s_{\ell-1,j}-1]}R_{i(j')},         \label{eq:lem:crossing:tower:dic:each:l:s}
%&\le \sum_{j'=s_{\ell,j}}^{s_{\ell-1,j}-1}f(T_3(j');T_4(j,\ell,j'))-\sum_{j'=s_{\ell,j}+1}^{s_{\ell-1,j}-1}R_{i(j')},         \label{eq:lem:crossing:tower:dic:each:l:s}
\end{align}
where the first inequality follows from Lemma \ref{lem:basic:tower:dic}, the second inequality follows from Lemma \ref{lem:horizontal:dic} with $a=s_{\ell,j}$, $b=s_{\ell-1,j}-1$, and the third inequality follows from Lemma \ref{lem:wac:linking} with $a=d=j'$, $b=j'+1$, and $c=s_{\ell-1,j}$ for any $j'\in [s_{\ell,j}+1:s_{\ell-1,j}-1]$. 
\iffalse
, and that
\begin{align}
T_3(j')&=\{ i(j'),i(j'+1),K_{[h_{j'}]}(j') \},     \nonumber    \\
T_4(j,\ell,j')&=\{ i(j'),i(s_{\ell-1,j}),K_{[h_{j'}]}(j') \}.    \nonumber
\end{align}
\fi
Similarly, one can show that 
\begin{align}
&F_{\ell}^t \le \sum_{j'\in [t_{\ell-1,j}:t_{\ell,j}-1]}f(T_3(j');T_5(j,\ell,j'))     \nonumber    \\
&\quad -\sum_{j'\in [t_{\ell-1,j}+1:t_{\ell,j}-1]}R_{i(j')}.          \label{eq:lem:crossing:tower:dic:each:l:t}
%&F_{\ell}^t \le \sum_{j'=t_{\ell-1,j}}^{t_{\ell,j}-1}f(T_3(j');T_5(j,\ell,j'))-\sum_{j'=t_{\ell-1,j}+1}^{t_{\ell,j}-1}R_{i(j')}.          \label{eq:lem:crossing:tower:dic:each:l:t}
\end{align}
\iffalse
where 
\begin{align}
T_5(j,\ell,j')=\{ i(j'),i(t_{\ell,j}),K_{[h_{j'}]}(j') \}.    \nonumber
\end{align}
\fi
By the submodularlity of $f(L;S)$ and Lemma \ref{lem:decoding:Bi:dic}, we have
\begin{align}
F_{\ell}&\ge R_{i(s_{\ell-1,j})}+R_{i(t_{\ell-1,j})}          \nonumber    \\
          &\quad +f([n]; i(s_{\ell,j}),i(s_{\ell-1,j}),i(t_{\ell-1,j}),i(t_{\ell,j}) ).   \label{eq:lem:crossing:tower:dic:each:l}
\end{align} 

Combining \eqref{eq:lem:crossing:tower:dic:each:l:s}-\eqref{eq:lem:crossing:tower:dic:each:l} and rearranging, we have
\begin{align}
&F_{\ell}^s+F_{\ell}^t-F_{\ell}+\sum_{j'=s_{\ell,j}+1}^{s_{\ell-1,j}}R_{i(j')}+\sum_{j'=t_{\ell-1,j}}^{t_{\ell,j}-1}R_{i(j')}     \nonumber    \\
&\le \sum_{j'\in [s_{\ell,j}:s_{\ell-1,j}-1]}f(T_3(j');T_4(j,\ell,j'))      \nonumber   \\
&\quad +\sum_{j'\in [t_{\ell-1,j}:t_{\ell,j}-1]}f(T_3(j');T_5(j,\ell,j'))     \nonumber    \\
%&\quad -\sum_{j'\in [s_{\ell,j}+1:s_{\ell-1,j}]}R_{i(j')}-\sum_{j'\in [t_{\ell-1,j}:t_{\ell,j}-1]}R_{i(j')}      \nonumber    \\
&\quad -f([n]; i(s_{\ell,j}),i(s_{\ell-1,j}),i(t_{\ell-1,j}),i(t_{\ell,j}) ).              \label{eq:lem:crossing:tower:dic:each:l:all}
\end{align}

Summing up \eqref{eq:lem:crossing:tower:dic:each:l:all} for all $\ell \in [2:h_j]$ yields
\begin{align}
&\sum_{\ell \in [2:h_j]}(F_{\ell}^s+F_{\ell}^t-F_{\ell})+\sum_{\ell \in [s_{h_j,j}+1:t_{h_j,j}-1]}R_{i(\ell)}         \nonumber      \\
%&\le \sum_{\ell \in [2:h_j]}( \sum_{j'=s_{\ell,j}}^{s_{\ell-1,j}-1}f(T_3,T_4)+\sum_{j'=t_{\ell-1,j}}^{t_{\ell,j}-1}f(T_3,T_5) )     \nonumber    \\
&\le \sum_{\ell \in [2:h_j]}\sum_{j'\in [s_{\ell,j}:s_{\ell-1,j}-1]}f(T_3(j');T_4(j,\ell,j'))     \nonumber    \\
&\quad +\sum_{\ell \in [2:h_j]}\sum_{j'\in [t_{\ell-1,j}:t_{\ell,j}-1]}f(T_3(j');T_5(j,\ell,j'))      \nonumber     \\
&\quad -\sum_{\ell \in [2:h_j]}f([n]; i(s_{\ell,j}),i(s_{\ell-1,j}),i(t_{\ell-1,j}),i(t_{\ell,j}) ).           \label{eq:lem:crossing:tower:dic:mid}
\end{align}

Finally, \eqref{eq:lem:crossing:tower:dic:goal:new} can be shown as follows, 
\begin{align}
&\sum_{\ell\in [h_j]}f([n]; i(s_{\ell,j}),i(t_{\ell,j}) )+\sum_{\ell \in [h_j]}R_{k_{\ell}(j)}    \nonumber    \\
&\quad +\sum_{\ell \in [2:h_j]}(F_{\ell}^s+F_{\ell}^t-F_{\ell})+\sum_{\ell \in [s_{h_j,j}+1:t_{h_j,j}-1]}R_{i(\ell)}         \nonumber      \\
&\le \sum_{\ell \in [2:h_j]}f([n]; i(s_{\ell,j}),i(t_{\ell,j}),i(s_{\ell-1,j}),i(t_{\ell-1,j}) )  \nonumber   \\
&\quad +f([n]; K_{[h_j]}(j),i(s_{h_j,j}),i(t_{h_j,j}) )                 \nonumber     \\
&\quad +\sum_{\ell \in [2:h_j]}\sum_{j'\in [s_{\ell,j}:s_{\ell-1,j}-1]}f(T_3(j');T_4(j,\ell,j'))     \nonumber    \\
&\quad +\sum_{\ell \in [2:h_j]}\sum_{j'\in [t_{\ell-1,j}:t_{\ell,j}-1]}f(T_3(j');T_5(j,\ell,j'))      \nonumber     \\
&\quad -\sum_{\ell \in [2:h_j]}f([n]; i(s_{\ell,j}),i(s_{\ell-1,j}),i(t_{\ell-1,j}),i(t_{\ell,j}) )          \nonumber     \\
%%
%&=f([n]; K_{[h_j]}(j),i(s_{h_j,j}),i(t_{h_j,j}) )           \nonumber     \\
%&\quad +\sum_{\ell \in [2:h_j]}\sum_{j'\in [s_{\ell,j}:s_{\ell-1,j}-1]}f(T_3(j');T_4(j,\ell,j'))     \nonumber    \\
%&\quad +\sum_{\ell \in [2:h_j]}\sum_{j'\in [t_{\ell-1,j}:t_{\ell,j}-1]}f(T_3(j');T_5(j,\ell,j'))      \nonumber     \\
%%
&\le f([n]; K_{[h_j]}(j),i(s_{h_j,j}),i(t_{h_j,j}) )           \nonumber     \\
&\quad +\sum_{\ell \in [2:h_j]}\sum_{j'\in [s_{\ell,j}:s_{\ell-1,j}-1]}\sum_{\substack{J\in N: J\cap T_3(j')\neq \emptyset, J\cap T_4(j,\ell,j')\neq \emptyset}}C_J            \nonumber    \\
&\quad +\sum_{\ell \in [2:h_j]}\sum_{j'\in [t_{\ell-1,j}:t_{\ell,j}-1]}\sum_{\substack{J\in N:J\cap T_3(j')\neq \emptyset, J\cap T_5(j,\ell,j')\neq \emptyset}}C_J,            \nonumber
\end{align}
where the first inequality follows from \eqref{eq:lem:crossing:tower:core3:dic} and \eqref{eq:lem:crossing:tower:dic:mid}, and the second inequality follows from property \eqref{eq:f:capacity:constraint} of $f(L;S)$. 
\end{IEEEproof}

%\vspace{5mm}
We show Theorem \ref{thm:gwac:dic}, i.e., \eqref{eq:gwac:dic:repeat}, in the following.
\begin{IEEEproof}
Consider the disjoint weighted alignment chain $i(1)\xleftrightarrow{\Ic,\Kcal } i(m+1)$ for a DIC problem. By Condition \ref{con:gwac:1} in Definition \ref{def:gwac:cic} and Lemma \ref{lem:decoding:Bi:dic}, \eqref{eq:wac:dic:terminals} also holds here. %we have
\iffalse
\begin{align}
R_{i_0}+R_{i_m}=f([n],\{ i(1),i(m+1) \}).           \label{eq:gwac:dic:terminals}
\end{align}
\fi

We first consider the special case when $|M\cup M'|=1$, which indicates that either $M$ or $M'$ is an empty set. 

If $M$ is an empty set, then $|M'|=1$. Hence, the disjoint weighted alignment chain reduces to a singleton weighted alignment chain of length $m=1$, %. It can be easily verified that for the special disjoint weighted alignment chains that are also singleton weighted alignment chains, 
where Theorem \ref{thm:gwac:dic} reduces to the already proved Theorem \ref{thm:wac:dic}. 

If $M'$ is an empty set, then $|M|=1$. Hence, the disjoint weighted alignment chain can be seen as a special crossing tower $\Xcal_j$ for which $M=\{ j \}$, $s_{h_j,j}=1$, $t_{h_j,j}=m+1$, and that $\{ i(s_{h_j,j}),i(t_{h_j,j}) \}$ is acyclic. In such case, 
\begin{align*}
T_2(j)&=\{ i(s_{h_j,j}),i(m+1),K_{[h_j]}(j) \}=T_1(j)|_{t_{h_j,j}=m+1}, 
\end{align*}
and hence \eqref{eq:gwac:dic:repeat} reduces to the following,
\begin{align}
R_{\rm sym} &\leq 
\frac{1}{|\Ic|+|\Kcal|}\big(\sum_{\substack{J\in N: J\cap T_1(j) \neq \emptyset}}C_J    \nonumber   \\
&\quad +\sum_{\ell \in [2:h_j]}( \sum_{j'\in [s_{\ell,j}:s_{\ell-1,j}-1]}\sum_{\substack{J\in N: J\cap T_3(j')\neq \emptyset,\\ J\cap T_4(j,\ell,j')\neq \emptyset}}C_J       \nonumber  \\ 
&\quad +\sum_{j'\in [t_{\ell-1,j}:t_{\ell,j}-1]}\sum_{\substack{J\in N: J\cap T_3(j')\neq \emptyset,\\ J\cap T_5(j,\ell,j')\neq \emptyset}}C_J )  \big).      \label{eq:gwac:dic:only:M:goal}
\end{align}
\iffalse
where 
\begin{align*}
T_1&=\{ i(s_{h_j,j}),i(t_{h_j,j}),K_{[h_j]}(j) \},   \nonumber    \\
T_2&=\{ i(s_{h_j,j}),i(m+1),K_{[h_j]}(j) \}=T_1|_{t_{h_j,j}=m+1}.      \nonumber     \\
\end{align*}
\fi

We have
\begin{align}
&(|\Ic|+|\Kcal|)R_{\rm sym}     \nonumber    \\
%%
%&=\sum_{j'\in G_j}\sum_{\ell\in [h_{j'}]}R_{k_{\ell}(j')}     \nonumber  \\
%&\quad +R_{i(s_{h_j,j})}+R_{i(t_{h_j,j})}+\sum_{\ell \in [s_{h_j,j}+1:t_{h_j,j}-1]}R_{i(\ell)}    \nonumber   \\
%%
&=\sum_{j'\in G_j}\sum_{\ell\in [h_{j'}]}R_{k_{\ell}(j')}     \nonumber  \\
&\quad +f([n]; i(s_{h_j,j}),i(t_{h_j,j}) )+\sum_{\ell \in [s_{h_j,j}+1:t_{h_j,j}-1]}R_{i(\ell)}    \nonumber   \\ 
&\le f([n]; K_{[h_j]}(j),i(s_{h_j,j}),i(t_{h_j,j}) )   \nonumber  \\
&\quad +\sum_{\ell \in [2:h_j]}\sum_{j' \in [s_{\ell,j}:s_{\ell-1,j}-1]}\sum_{\substack{J\in N: J\cap T_3(j')\neq \emptyset, \\ J\cap T_4(j,\ell,j')\neq \emptyset}}C_J  \nonumber  \\
&\quad +\sum_{\ell \in [2:h_j]}\sum_{j' \in [t_{\ell-1,j}:t_{\ell,j}-1]}\sum_{\substack{J\in N: J\cap T_3(j')\neq \emptyset, \\ J\cap T_5(j,\ell,j')\neq \emptyset}}C_J    \nonumber   \\
&\le \sum_{\substack{J\in N: J\cap T_1(j) \neq \emptyset}}C_J   \nonumber  \\
&\quad +\sum_{\ell \in [2:h_j]}\sum_{j' \in [s_{\ell,j}:s_{\ell-1,j}-1]}\sum_{\substack{J\in N: J\cap T_3(j')\neq \emptyset, \\ J\cap T_4(j,\ell,j')\neq \emptyset}}C_J  \nonumber  \\
&\quad +\sum_{\ell \in [2:h_j]}\sum_{j' \in [t_{\ell-1,j}:t_{\ell,j}-1]}\sum_{\substack{J\in N: J\cap T_3(j')\neq \emptyset, \\ J\cap T_5(j,\ell,j')\neq \emptyset}}C_J,
\end{align}
where the equality follows from \eqref{eq:wac:dic:terminals}, the first inequality follows from Lemma \ref{lem:x:tower:dic}, and the second inequality follows from property \eqref{eq:f:capacity:constraint} of $f(L;S)$. The above inequality directly leads to \eqref{eq:gwac:dic:only:M:goal} and thus completes the proof. 

So far we have shown that Theorem \ref{thm:gwac:dic} holds when $|M\cup M'|=1$. From now on, we assume that $|M\cup M'|\ge 2$. 

As any disjoint weighted alignment chain can be seen as a concatenation of the crossing towers $\Xcal_j$, $j\in M$, and the basic tower $\Bc_{j'}$, $j'\in M'$, by Lemmas \ref{lem:basic:tower:dic} and \ref{lem:x:tower:dic}, we have
\begin{align}
&\sum_{j\in M}\big( \sum_{j'\in G_j}\sum_{\ell\in [h_{j'}]}R_{k_{\ell}(j')}     \nonumber   \\
&\quad +f([n]; i(s_{h_j,j}),i(t_{h_j,j}) )+\sum_{j' \in [s_{h_j,j}+1:t_{h_j,j}-1]}R_{i(j')}       \big)   \nonumber   \\
&\quad +\sum_{j' \in M'}\big( \sum_{\ell\in [h_{j'}]}R_{k_{\ell}(j')}+f([n]; i(j'),i(j'+1) ) \big)      \nonumber     \\
%%%
&\le \sum_{j\in M}\big( f([n]; K_{[h_j]}(j),i(s_{h_j,j}),i(t_{h_j,j}) )           \nonumber     \\
&\quad +\sum_{\ell \in [2:h_j]}\sum_{j' \in [s_{\ell,j}:s_{\ell-1,j}-1]}\sum_{\substack{J\in N: J\cap T_3(j')\neq \emptyset, \\ J\cap T_4(j,\ell,j')\neq \emptyset}}C_J  \nonumber  \\
&\quad +\sum_{\ell \in [2:h_j]}\sum_{j' \in [t_{\ell-1,j}:t_{\ell,j}-1]}\sum_{\substack{J\in N: J\cap T_3(j')\neq \emptyset, \\ J\cap T_5(j,\ell,j')\neq \emptyset}}C_J \big)            \nonumber       \\
&\quad +\sum_{j'\in M'}f([n]; i(j'),i(j'+1),K_{[h_{j'}]}(j') ).    \label{eq:gwac:dic:final}
\end{align}

%\textcolor{red}{tbc}

Same as what we have done in the proof of Thoerem \ref{thm:gwac:cic} in Appendex \ref{sec:modular:gwac:cic}, we can reindex the elements of the set $M\cup M'$ as $j_1<j_2<\cdots<j_{|M\cup M'|}$, and \eqref{eq:gwac:cic:order:left:end}-\eqref{eq:gwac:cic:order:dula} hold, which are repeated below for easier reference. 
\begin{align}
&s_{h_{j_1},j_1}=1,         \label{eq:gwac:dic:order:left:end}   \\
&t_{h_{j_{|M\cup M'|}},j_{|M\cup M'|}}=m+1,             \label{eq:gwac:dic:order:right:end}    \\
&s_{h_{j_{p}},j_{p}}=t_{h_{j_{p-1}},j_{p-1}},  \qquad \quad \forall p \in [2:|M\cup M'|].   \label{eq:gwac:dic:order:dula}
\end{align}

For any $p\in [|M\cup M'|-1]$, by the submodularity of $f(L;S)$ and Lemma \ref{lem:decoding:Bi:dic}, we have
\begin{align}
&f([n]; i(m+1),i(t_{h_{j_{p}},j_{p}}) )+f([n]; i(t_{h_{j_{p}},j_{p}}),i(s_{h_{j_{p}},j_{p}}) )          \nonumber    \\
&\ge R_{i(t_{h_{j_{p}},j_{p}})}+f([n]; i(m+1),i(t_{h_{j_{p}},j_{p}}),i(s_{h_{j_{p}},j_{p}}) )             \nonumber    \\
&=R_{i(t_{h_{j_{p}},j_{p}})}+f([n]; i(m+1),i(s_{h_{j_{p}},j_{p}}) )        \nonumber   \\
&\quad +f([n]; i(m+1),i(t_{h_{j_{p}},j_{p}}),i(s_{h_{j_{p}},j_{p}}) )             \nonumber    \\
&\quad -f([n]; i(m+1),i(s_{h_{j_{p}},j_{p}}) )).     \label{eq:gwac:dic:order:each:l}
\end{align}

Given the fact that $j_p$, $p\in [|M\cup M'|]$ is a reindexing of $j$, $j\in M\cup M'$, as well as \eqref{eq:gwac:dic:order:left:end}-\eqref{eq:gwac:dic:order:dula}, summing up \eqref{eq:gwac:dic:order:each:l} for all $p \in [|M\cup M'|-1]$ yields
\begin{align}
&\sum_{j\in M\cup M'}f([n]; i(s_{h_j,j}),i(t_{h_j,j}) )        \nonumber     \\
%&\sum_{p \in [|M\cup M'|]}f([n]; i(s_{h_{j_p},j_p}),i(t_{h_{j_p},j_p}) )        \nonumber     \\
%%
&\ge \sum_{j' \in (\bigcup_{j\in M\cup M'}\{ s_{h_j,j},t_{h_j,j} \})\setminus \{ 1,m+1 \}}R_{i(j')}     \nonumber   \\
&\quad +f([n]; i(m+1),i(1) )    \nonumber    \\
&\quad +\sum_{p \in [|M\cup M'|-1]}f([n];  i(m+1),i(t_{h_{j_{p}},j_{p}}),i(s_{h_{j_{p}},j_{p}}) )    \nonumber    \\
&\quad -\sum_{p \in [|M\cup M'|-1]}f([n]; i(m+1),i(s_{h_{j_{p}},j_{p}}) )    \nonumber   \\
&= \sum_{j' \in \bigcup_{j\in M\cup M'}\{ s_{h_j,j},t_{h_j,j} \}}R_{i(j')}    \nonumber   \\
&\quad +\sum_{p \in [|M\cup M'|-1]}f([n];  i(m+1),i(t_{h_{j_{p}},j_{p}}),i(s_{h_{j_{p}},j_{p}}) )    \nonumber    \\
&\quad -\sum_{p \in [|M\cup M'|-1]}f([n]; i(m+1),i(s_{h_{j_{p}},j_{p}}) ),  \label{eq:gwac:dic:final:LHS:mid} 
\end{align}
where the equality is due to \eqref{eq:wac:dic:terminals}.

The LHS of \eqref{eq:gwac:dic:final} can be bounded as,
\begin{align}
&LHS  \nonumber  \\
=&\sum_{j\in M}\sum_{j'\in G_j}\sum_{\ell\in [h_{j'}]}R_{k_{\ell}(j')}+\sum_{j' \in M'}\sum_{\ell\in [h_{j'}]}R_{k_{\ell}(j')}   \nonumber    \\
&+\sum_{j\in M}\sum_{j' \in [s_{h_j,j}+1:t_{h_j,j}-1]}R_{i(j')}                  \nonumber     \\
&+\sum_{j\in M\cup M'}f([n]; i(s_{h_j,j}),i(t_{h_j,j}) )      \nonumber     \\
\ge&\sum_{j\in [m]}\sum_{\ell\in [h_j]}R_{k_{\ell}(j)}+\sum_{j\in M}\sum_{j' \in [s_{h_j,j}+1:t_{h_j,j}-1]}R_{i(j')}       \nonumber     \\
&+\sum_{j' \in \bigcup_{j\in M\cup M'}\{ s_{h_j,j},t_{h_j,j} \}}R_{i(j')}    \nonumber   \\
&+\sum_{p \in [|M\cup M'|-1]}f([n];  i(m+1),i(t_{h_{j_{p}},j_{p}}),i(s_{h_{j_{p}},j_{p}}) )    \nonumber    \\
&-\sum_{p \in [|M\cup M'|-1]}f([n]; i(m+1),i(s_{h_{j_{p}},j_{p}}) )   \nonumber   \\
%%%
=&(|\Kcal|+|\Ic|)R_{\rm sym}    \nonumber    \\
&+\sum_{p \in [|M\cup M'|]}f([n];  i(m+1),i(t_{h_{j_{p}},j_{p}}),i(s_{h_{j_{p}},j_{p}}) )    \nonumber    \\
&-\sum_{p \in [|M\cup M'|]}f([n]; i(m+1),i(s_{h_{j_{p}},j_{p}}) )   \nonumber   \\
%%%
=&(|\Kcal|+|\Ic|)R_{\rm sym}    \nonumber    \\
&+\sum_{j \in M\cup M'}f([n];  i(m+1),i(t_{h_{j},j}),i(s_{h_{j},j}) )    \nonumber    \\
&-\sum_{j \in M\cup M'}f([n]; i(m+1),i(s_{h_{j},j}) ),   %\nonumber   \\
%%%
%\ge&\sum_{\ell \in [|M\cup M'|-1]}\big( f([n],\{ K_{v_{\ell},[w_{v_{\ell}}]},i_{p_{v_{\ell},w_{v_{\ell}}}},i_{q_{v_{\ell},w_{v_{\ell}}}} \})-   \nonumber   \\
%&f(\{ K_{v_{\ell},[w_{v_{\ell}}]},i_{p_{v_{\ell},w_{v_{\ell}}}},i_{q_{v_{\ell},w_{v_{\ell}}}} \},\{ K_{v_{\ell},[w_{v_{\ell}}]},i_m,i_{p_{v_{\ell},w_{v_{\ell}}}} \}) \big)        \nonumber  \\
%&+(|\Kcal|+|\Ic|)R_{\rm sym}    
\label{eq:gwac:dic:final:LHS}
\end{align}
where the first equality follows from simply rearranging the terms of the LHS of \eqref{eq:gwac:dic:final}, the inequality follows from \eqref{eq:gwac:dic:final:LHS:mid}, the second equality follows from 
%the fact that $f([n]; i(m+1),i(t_{h_{j_{p}},j_{p}}),i(s_{h_{j_{p}},j_{p}}) )|_{p=|M\cup M'|}=$, which is due to 
\eqref{eq:gwac:dic:order:right:end}, and the third equality follows from the fact that $j_p$, $p\in [|M\cup M'|]$ is a reindexing of $j$, $j\in M\cup M'$. 

\iffalse
the fact that for any $\ell \in [|M\cup M'|-1]$ we have
\begin{align*}
&f([n],\{  i_{m},i_{q_{v_{\ell},w_{v_{\ell}}}},i_{p_{v_{\ell},w_{v_{\ell}}}} \})-f([n],\{ i_{m},i_{p_{v_{\ell},w_{v_{\ell}}}} \})    \\
&\ge f([n],\{ K_{v_{\ell},[w_{v_{\ell}}]},i_{p_{v_{\ell},w_{v_{\ell}}}},i_{q_{v_{\ell},w_{v_{\ell}}}} \})-      \\
&\quad f(\{ K_{v_{\ell},[w_{v_{\ell}}]},i_{p_{v_{\ell},w_{v_{\ell}}}},i_{q_{v_{\ell},w_{v_{\ell}}}} \},\{ K_{v_{\ell},[w_{v_{\ell}}]},i_m,i_{p_{v_{\ell},w_{v_{\ell}}}} \}),
\end{align*}
which can be shown using the submodularity and monotonicity of $f(L,S)$, together with \eqref{eq:linking}, similar to proving \eqref{eq:gwac:dic:special}.
\fi

For the RHS of \eqref{eq:gwac:dic:final}, simply rearranging the terms we have
\begin{align}
&RHS   \nonumber   \\
%%%
=&\sum_{j\in M\cup M'}\big( f([n]; K_{[h_j]}(j),i(s_{h_j,j}),i(t_{h_j,j}) )         \nonumber    \\
&+\sum_{j\in M}\big( \sum_{\ell \in [2:h_j]}\sum_{j' \in [s_{\ell,j}:s_{\ell-1,j}-1]}\sum_{\substack{J\in N: J\cap T_3(j')\neq \emptyset, \\ J\cap T_4(j,\ell,j')\neq \emptyset}}C_J  \nonumber  \\
&\quad +\sum_{\ell \in [2:h_j]}\sum_{j' \in [t_{\ell-1,j}:t_{\ell,j}-1]}\sum_{\substack{J\in N: J\cap T_3(j')\neq \emptyset, \\ J\cap T_5(j,\ell,j')\neq \emptyset}}C_J \big).     %            \nonumber       \\
%%%
%=&\sum_{\ell \in [|M\cup M'|]}f([n],\{ K_{[h_{v_{\ell}}]}(v_{\ell}),i(s_{h_{v_{\ell}},v_{\ell}}),i(t_{h_{v_{\ell}},v_{\ell}}) \})     \nonumber   \\
%&+\sum_{j\in M}\big( \sum_{\ell \in [2:h_j]}\sum_{j'=s_{\ell,j}}^{s_{\ell-1,j}-1}\sum_{J\subseteq [n]:J\cap T_3\neq \emptyset,J\cap T_4\neq \emptyset}C_J            \nonumber    \\
%&+\sum_{\ell \in [2:h_j]}\sum_{j'=t_{\ell-1,j}}^{t_{\ell,j}-1}\sum_{J\subseteq [n]:J\cap T_3\neq \emptyset,J\cap T_5\neq \emptyset}C_J \big),     
\label{eq:gwac:dic:final:RHS}
\end{align}
%where the first equality follows from simply rearranging the terms of the RHS of \eqref{eq:gwac:dic:final}

\iffalse
Set  
\begin{align*}
T_{v,1}&=\{ i(s_{h_{v_{\ell}},{v_{\ell}}}),i(t_{h_{v_{\ell}},{v_{\ell}}}),K_{[h_{v_{\ell}}]}(v_{\ell}) \}     \\
T_{v,2}&=\{ i(s_{h_{v_{\ell}},v_{\ell}}),i(m+1),K_{[h_{v_{\ell}}]}(v_{\ell}) \}.
\end{align*} 
\fi

Given \eqref{eq:gwac:dic:final}, \eqref{eq:gwac:dic:final:LHS}, and \eqref{eq:gwac:dic:final:RHS}, we can conclude that %$R_{\rm sym}\le \frac{m}{|\Kcal|+|\Ic|}.$
%\iffalse
\begin{align}
&(|\Kcal|+|\Ic|)R_{\rm sym}        \nonumber   \\
&-\sum_{j\in M}\big( \sum_{\ell \in [2:h_j]}\sum_{j' \in [s_{\ell,j}:s_{\ell-1,j}-1]}\sum_{\substack{J\in N: J\cap T_3(j')\neq \emptyset, \\ J\cap T_4(j,\ell,j')\neq \emptyset}}C_J  \nonumber  \\
&\quad +\sum_{\ell \in [2:h_j]}\sum_{j' \in [t_{\ell-1,j}:t_{\ell,j}-1]}\sum_{\substack{J\in N: J\cap T_3(j')\neq \emptyset, \\ J\cap T_5(j,\ell,j')\neq \emptyset}}C_J \big)   \nonumber   \\
%%%
\le &\sum_{j\in M\cup M'}\big( f([n]; K_{[h_j]}(j),i(s_{h_j,j}),i(t_{h_j,j}) )     \nonumber    \\
&-\sum_{j \in M\cup M'}f([n];  i(m+1),i(t_{h_{j},j}),i(s_{h_{j},j}) )    \nonumber    \\
&+\sum_{j \in M\cup M'}f([n]; i(m+1),i(s_{h_{j},j}) )  \nonumber   \\
%%%
%\le &\sum_{\ell \in [|M\cup M'|]}f([n],\{ K_{[h_{v_{\ell}}]}(v_{\ell}),i(s_{h_{v_{\ell}},v_{\ell}}),i(t_{h_{v_{\ell}},v_{\ell}}) \})      \nonumber   \\
%&-\sum_{\ell \in [|M\cup M'|]}f([n],\{  i(m+1),i(t_{h_{v_{\ell}},v_{\ell}}),i(s_{h_{v_{\ell}},v_{\ell}}) \})    \nonumber    \\
%&+\sum_{\ell \in [|M\cup M'|]}f([n],\{ i(m+1),i(s_{h_{v_{\ell}},v_{\ell}}) \}))   \nonumber  \\
%%%
%\le &\sum_{\ell \in [|M\cup M'|]}f(T_{v,1},T_{v,2}),   \nonumber   \\
%%%
\le &\sum_{j \in M\cup M'}f(T_{1}(j);T_{2}(j))   \nonumber   \\
%%%
\le&\sum_{j\in M\cup M'}\sum_{J\in N: J\cap T_1(j)\neq \emptyset, J\cap T_2(j)\neq \emptyset}C_J,   \label{eq:gwac:dic:final:final}
\end{align}
where the second inequality follows from Lemma \ref{lem:wac:linking} with $a=s_{h_j,j}$, $b=t_{h_j,j}$, $c=m+1$, and $d=j$ for any $j\in M\cup M'$, 
\iffalse
and that 
\begin{align*}
T_1&=\{ i(s_{h_j,j}),i(t_{h_j,j}),K_{[h_j]}(j) \},   \\
T_2&=\{ i(s_{h_j,j}),i(m+1),K_{[h_j]}(j) \},
\end{align*} 
\fi
and the last inequality follows from property \eqref{eq:f:capacity:constraint} of $f(L;S)$.

\iffalse
where the equality is due to \eqref{eq:linking} as well as the fact that
\begin{align*}
&f([n],\{ K_{v_{\ell},[w_{v_{\ell}}]},i_{p_{v_{\ell},w_{v_{\ell}}}},i_{q_{v_{\ell},w_{v_{\ell}}}} \})|_{\ell=|M\cup M'|}  \\
&=f([n],\{ K_{v_{|M\cup M'|},[w_{v_{|M\cup M'|}}]},i_{p_{v_{|M\cup M'|},w_{v_{|M\cup M'|}}}},i_m \}),
\end{align*}
and the last inequality is due to \eqref{eq:capacity:constraint}.
\fi

Rearranging and simplifying \eqref{eq:gwac:dic:final:final} yields \eqref{eq:gwac:dic:repeat} and thus completes the proof. 
\end{IEEEproof}

%%%%%%%%%%%%%%%%%%%%%%%%%%%%%%%%%%%%%%%%%%%%
%%%%%%%%%%%%%%%%%%%%%%%%%%%%%%%%%%%%%%%%%%%%
%%%%%%%%%%%%%%%%%%%%%%%%%%%%%%%%%%%%%%%%%%%%
%NON-SHANNON_INEQUALITY EXAMPLE

%\vspace{2mm}
\section{Proof of \eqref{eq:zy:n9:goal}} \label{sec:proof:zy}

\begin{IEEEproof}
For the problem described in Example \ref{exm:zy}, we show that \eqref{eq:zy:n9:goal} holds by showing that 
\begin{align}
&11+g(1,2,3)+g(1,2,4)  \nonumber   \\
&\quad \ge 6R_1+6R_2+3R_3+3R_4+4R_5+4R_6   \nonumber  \\
&\quad \quad +2R_7+4R_8+4R_9.      \label{eq:goal:shannon}
\end{align}
and that
\begin{align}
&14-g(1,2,3)-g(1,2,4)  \nonumber   \\
&\quad \ge 2R_1+2R_2+5R_3+5R_4+4R_5+4R_6   \nonumber  \\
&\quad \quad +6R_7+4R_8+4R_9.        \label{eq:goal:nonshannon}
\end{align}
in the following.

We first show that \eqref{eq:goal:shannon} holds as follows.

First of all, it can be verified the following inequalities hold
\begin{align}
g(1,3,4)&\ge g(1,4)+g(1,3)+R_6+R_9-1,     \label{eq:zy:mid:1}    \\
g(2,3,4)&\ge g(2,4)+g(2,3)+R_5+R_8-1,     \label{eq:zy:mid:2}    \\
g(1,3,5)&\ge g(3,5)+g(1,5)+R_2+R_8-1,     \label{eq:zy:mid:3}    \\
g(1,4,8)&\ge g(4,8)+g(1,8)+R_2+R_5-1,     \label{eq:zy:mid:4}    \\
g(1,2,3)&\ge g(2,3)+g(1,5)+R_2+R_8-1,     \label{eq:zy:mid:5}    \\
g(1,2,4)&\ge g(2,4)+g(1,8)+R_2+R_5-1,     \label{eq:zy:mid:6}
\end{align}
among which we only give detailed derivations for the first one, \eqref{eq:zy:mid:1}, as shown in the following, while all the others can be obtained via similar steps. Consider
\begin{align}
g(1,3,4)+1-R_9&\ge g(1,3,4)+g(1,4,6,9)-R_9   \nonumber   \\
                    &= g(1,3,4)+g(1,4,6)   \nonumber  \\
                    &\ge g(1,4)+g(1,3,4,6)    \nonumber   \\
                    &\ge g(1,4)+g(1,3,6)   \nonumber   \\
                    &=g(1,4)+g(1,3)+R_6,   \label{eq:zy:mid:derivation:1}
\end{align}
where the first inequality follows from the fact that $g(S)\le 1,\forall S\subseteq [n]$, the first equality follows from Lemma \ref{lem:decoding:Bi} with $\{1,4,6\} \subseteq B_9$, the second inequality follows from the submodularity of $g(S)$, the third inequality follows from the monotonicity of $g(S)$, and the second equality follows from Lemma \ref{lem:decoding:Bi} with $\{1,3\} \subseteq B_6$. Rearranging %the above inequality leads to \eqref{eq:zy:mid:1}. 
\eqref{eq:zy:mid:derivation:1} leads to \eqref{eq:zy:mid:1}.

Next, it can be verified that according to Lemma \ref{lem:decoding:Bi}, as well as the submodularity and the monotonicity of $g(S)$, we have the following inequalities,
\begin{align}
&g(2,3)+g(1,3)+g(2,4)+g(1,4)   \nonumber  \\
&\quad \ge R_3+R_4+g(1,2,3,4)+R_1+R_2,   \label{eq:zy:1}    \\
&g(1,2,3,4)+g(1,3,4,5,8)   \nonumber   \\
&\quad \ge g(1,3,4)+R_2+R_5+R_8,    \label{eq:zy:2}   \\
&g(1,3,4,5)+g(1,3,4,8)   \nonumber  \\
&\quad \ge g(1,3,4)+g(1,3,4,5,8),    \label{eq:zy:4}    \\
&2g(3,4)+g(3,5)+g(4,8)    \nonumber   \\
&\quad \ge R_3+R_4+g(3,4,5)+g(3,4,8),    \label{eq:zy:6}    \\
&g(3,4,5)+g(1,4,5)+g(3,4,8)+g(1,3,8)     \nonumber     \\
&\ge R_4+R_5+g(1,3,4,5)+R_3+R_8+g(1,3,4,8),     \label{eq:zy:7}     \\
%%%%%%%%%%%%%
&g(1,3)+g(1,5)+g(1,4)+g(1,8)   \nonumber   \\
&\quad \ge g(1,3,5)+g(1,4,8)+2R_1,   \label{eq:zy:11}    \\
&3\ge R_1+2R_6+2R_9+g(1,3)+g(1,4),    \label{eq:zy:14}    \\
&2\ge 2R_7+2g(3,4).  \label{eq:zy:17}
\end{align}

By \eqref{eq:zy:mid:1}, we have
\begin{align}
2+2g(1,3,4)\ge 2(g(1,4)+g(1,3)+R_6+R_9).  \label{eq:zy:15}
\end{align}

By \eqref{eq:zy:mid:3} and \eqref{eq:zy:mid:4}, and the submodularity of $g(S)$, as well as Lemma \ref{lem:decoding:Bi}, it can be verified that
\begin{align}
&g(1,3,5)+g(1,4)+g(1,4,8)+g(1,3)     \nonumber  \\
&\quad \ge g(3,5)+g(1,4,5)+g(4,8)+g(1,3,8)   \nonumber   \\
&\quad \quad +2R_1+2R_2+R_5+R_8-2.             \label{eq:zy:9}
\end{align}

Summing up \eqref{eq:zy:1}-\eqref{eq:zy:9} and simplifying, we have
\begin{align}
&9+g(2,3)+g(2,4)+g(1,5)+g(1,8)   \nonumber   \\
&\quad \ge 6R_1+4R_2+3R_3+3R_4+3R_5+4R_6   \nonumber  \\
&\quad \quad +2R_7+3R_8+4R_9.   \label{eq:zy:19}
\end{align}

Then, adding \eqref{eq:zy:19}, \eqref{eq:zy:mid:5} and \eqref{eq:zy:mid:6} and simplifying, we obtain \eqref{eq:goal:shannon}. 

It remains to show that \eqref{eq:goal:nonshannon} holds, and for that purpose we will use the Zhang-Yeung non-Shannon-type information inequality \cite{zhang1998characterization}, stated as follows,
\begin{align}
&3H(\Ac,\Cc)+3H(\Ac,\Dc)+3H(\Cc,\Dc)+H(\Bc,\Cc)+H(\Bc,\Dc)   \nonumber  \\
&\quad \ge 2H(\Cc)+2H(\Dc)+H(\Ac,\Bc)+H(\Ac)   \nonumber   \\
&\quad \quad +H(\Bc,\Cc,\Dc)+4H(\Ac,\Cc,\Dc),    \label{eq:zy:zy}
\end{align}
where $\Ac,\Bc,\Cc,\Dc$ each denotes an arbitrary subset of the set of all random variables for the problem, $\{ Y_{[n]}, X_1,X_2,\cdots,X_n \}$. 
Set 
\begin{align}
\begin{array}{cc}
\Ac=\{Y_{[n]}\} \cup \Xv_{\{1,2,4\}^c}, & \Bc=\{Y_{[n]}\} \cup \Xv_{\{1,2,3\}^c},   \\
\Cc=\{Y_{[n]}\} \cup \Xv_{\{2,3,4\}^c}, & \Dc=\{Y_{[n]}\} \cup \Xv_{\{1,3,4\}^c}.
\end{array}  \label{eq:zy:set:abcd}
\end{align}
By \eqref{eq:zy:zy} and message independence, as well as the definition of the set function $g(S)$, we have
\begin{align}
&3g(2,4)+3g(1,4)+3g(3,4)+g(2,3)+g(1,3)\ge 2g(2,3,4)      \nonumber    \\
&+2g(1,3,4)+g(1,2)+g(1,2,4)+g(3)+4g(4).     \label{eq:zy:zy:1}
\end{align}
Swap $\Ac$ and $\Bc$ in \eqref{eq:zy:set:abcd}, but keep $\Cc$ and $\Dc$ unchanged. By \eqref{eq:zy:zy} and message independence, as well as the definition of the set function $g(S)$, we have 
\begin{align}
&3g(2,3)+3g(1,3)+3g(3,4)+g(2,4)+g(1,4)\ge 2g(2,3,4)      \nonumber    \\
&+2g(1,3,4)+g(1,2)+g(1,2,3)+g(4)+4g(3).     \label{eq:zy:zy:2}
\end{align}

Adding \eqref{eq:zy:zy:1} and \eqref{eq:zy:zy:2}, we obtain
\begin{align}
&4g(2,3)+4g(1,3)+6g(3,4)+4g(2,4)+4g(1,4)      \nonumber    \\
&\quad \ge 4g(2,3,4)+4g(1,3,4)+2g(1,2)+5g(3)+5g(4)    \nonumber    \\
&\quad \quad +g(1,2,3)+g(1,2,4)     \nonumber   \\
&\quad \ge 4g(2,3,4)+4g(1,3,4)+2R_1+2R_2+5R_3+5R_4    \nonumber   \\
&\quad \quad +g(1,2,3)+g(1,2,4),     \label{eq:zy:zy:3}
\end{align}
where the second inequality follows from Lemma \ref{lem:decoding:Bi} with $\{1\} \subseteq B_2$. 

Therefore, we have
\begin{align}
6&\ge 6g(3,4,7)   \nonumber   \\
 &=6R_7+6g(3,4)   \nonumber   \\
 &\ge 6R_7+2R_1+2R_2+5R_3+5R_4    \nonumber   \\
 &\quad +4\big( g(2,3,4)-g(2,4)-g(2,3) \big)   \nonumber    \\
 &\quad +4\big( g(1,3,4)-g(1,4)-g(1,3) \big)   \nonumber    \\
 &\quad +g(1,2,3)+g(1,2,4)    \nonumber   \\
 &\ge 6R_7+2R_1+2R_2+5R_3+5R_4    \nonumber   \\
 &\quad +4(R_5+R_8-1)+4(R_6+R_9-1)  \nonumber   \\
 &\quad +g(1,2,3)+g(1,2,4),    \label{eq:zy:result}
\end{align}
where the first inequality follows from that $g(S)\le 1,\forall S\subseteq [n]$, the equality follows from Lemma \ref{lem:decoding:Bi} with $\{3,4\} \subseteq B_7$, the second inequality follows from \eqref{eq:zy:zy:3}, and the last inequality follows from \eqref{eq:zy:mid:1} and \eqref{eq:zy:mid:2}. 

We can see that rearranging \eqref{eq:zy:result} leads to \eqref{eq:goal:nonshannon}. 

Now that both \eqref{eq:goal:shannon} and \eqref{eq:goal:nonshannon} have been proved, by adding them together and dividing both sides by $8$, we conclude that
\begin{align}
\sum_{i\in [n]}R_i\le \frac{25}{8},
\end{align}
which completes the proof.
\end{IEEEproof}

\section*{Acknowledgement}\label{sec:acknowledgement}
%\bigskip

%\textit{Acknowledgement:} 
The authors would like to thank Young-Han Kim from UCSD for many fruitful discussions.

%%%%%%%%%%%%%%%%%%%%%%%%%%%

\bibliographystyle{IEEEtran}
\bibliography{references} 

\end{document}